\documentclass{aa}

\usepackage{graphicx}
\usepackage{txfonts}

\usepackage{amsmath}    % Advanced maths commands
\usepackage{color}
\usepackage{xcolor}
\usepackage{colortbl}
\usepackage{nicefrac}

\usepackage{hyperref}
\newcolumntype{$}{>{\global\let\currentrowstyle\relax}}
\newcolumntype{^}{>{\currentrowstyle}}
\newcommand{\rowstyle}[1]{\gdef\currentrowstyle{#1}%
  #1\ignorespaces
}

\newcommand{\teh}{\textsc{T-800}}

\newcommand{\darwin}{\textsc{Darwin}}
\newcommand{\exomol}{\textsc{ExoMol}}

\newcommand{\rextf}{\ensuremath{r^{\text{ext}}_{\text{final}}}}

\newcommand{\radd}{\ensuremath{r_{\text{d}}}}

\newcommand{\FDi}{\ensuremath{\mathfrak{F}_{\text{D}}^{\infty}}}
\newcommand{\FDk}{\ensuremath{\mathfrak{F}_{\text{D},k}}}
\newcommand{\FDki}{\ensuremath{\mathfrak{F}_{\text{D},k}^{\infty}}}

\newcommand{\cs}{\ensuremath{c_{\text{s}}}}

\newcommand{\vD}{\ensuremath{v_{\text{D}}}}

\newcommand{\rhog}{\ensuremath{\rho_{\text{g}}}}
\newcommand\Kp[1]{\ensuremath{K_{\text{p}}\left({#1}\right)}}
\newcommand\Kps[2]{\ensuremath{K_{\text{p}}^{#1}\left({#2}\right)}}

\newcommand{\vzeta}{\ensuremath{v_{\zeta}}}
\newcommand{\ddg}{\ensuremath{\delta_{\text{dg}}}}
\newcommand{\ddgk}{\ensuremath{\delta_{\text{dg},k}}}

\newcommand{\vDeq}{\ensuremath{\mathring{v}_{\text{D}}}}
\newcommand{\vDinf}{\ensuremath{v_{\text{D},\infty}}}

\newcommand{\tauGi}{\ensuremath{\tau^{-1}_{\text{gr}}}}
\newcommand{\tauns}{\ensuremath{\tau^{-1}_{\text{ns}}}}
\newcommand{\taudc}{\ensuremath{\tau^{-1}_{\text{dc}}}}

\newcommand{\rhod}{\ensuremath{\rho_{\text{d}}}}
\newcommand{\rhodk}{\ensuremath{\rho_{\text{d},k}}}
\newcommand{\rhoint}[1]{\ensuremath{\rho_{\text{m},#1}}}
\newcommand{\nds}[1]{\ensuremath{n_{\text{d},#1}}}

\newcommand{\Td}{\ensuremath{T_{\text{d}}}}
\newcommand{\Tg}{\ensuremath{T_{\text{g}}}}
\newcommand{\Tr}{\ensuremath{T_{\text{r}}}}
\newcommand{\fdrag}{\ensuremath{f_{\text{drag}}}}
\newcommand{\fradd}{\ensuremath{f_{\text{rad,d}}}}
\newcommand{\fgravd}{\ensuremath{f_{\text{grav,d}}}}

\newcommand{\kappaH}{\ensuremath{\kappa_{\text{H}}}}

\newcommand{\kappaR}{\ensuremath{\kappa_{\text{R}}}}

\newcommand{\chiH}{\ensuremath{\chi_{\text{H}}}}

\newcommand{\chiR}{\ensuremath{\chi_{\text{R}}}}

\newcommand{\Qnuabs}{\ensuremath{Q_{\text{abs},\nu}}}

\newcommand{\Qnuabspr}{\ensuremath{Q_{\text{abs},\nu}(\text{pr})}}

\newcommand{\Qnuext}{\ensuremath{Q_{\text{ext},\nu}}}
\newcommand{\Qnusca}{\ensuremath{Q_{\text{sca},\nu}}}

\newcommand{\cost}{\ensuremath{\langle\cos\theta\rangle}}
\newcommand{\kB}{\ensuremath{k_{\text{B}}}}

\newcommand{\fedd}{\ensuremath{f_{\text{Edd}}}}
\newcommand{\kms}{\ensuremath{\,\mbox{km}\,\mbox{s}^{-1}}}

\newcommand{\drhog}{\ensuremath{\ddg\FDi}}
\newcommand{\drhogf}{\ensuremath{\displaystyle\ddg\FDi}}

\newcommand{\uinf}{\ensuremath{u_{\infty}}}

\newcommand{\vinfk}{\ensuremath{v_{\infty,k}}}

\newcommand{\fconde}[1]{\ensuremath{f_{\text{cond}}^{(\text{#1})}}}

\newcommand{\vdrinf}{\ensuremath{v_{\text{D},\infty}}}

\newcommand{\teff}{\ensuremath{T_{\text{eff}}}}
\newcommand{\deltauvelp}{\ensuremath{\Delta\,u_{\text{p}}}}
\newcommand{\mdotu}{\ensuremath{\text{M}_{\sun}\,\text{yr}^{-1}}}

\newcommand{\mdot}{\ensuremath{\dot{M}}}
\newcommand{\mmdot}{\ensuremath{\langle\mdot\rangle}}
\newcommand{\dmdot}{\ensuremath{\dot{M}_{\text{d}}}}

\newcommand{\mmdmdot}{\ensuremath{\langle\dmdot/\mdot\rangle}}
\newcommand{\muinf}{\ensuremath{\langle u_{\infty}\rangle}}
\newcommand{\mfcondsi}{\ensuremath{\langle f_{\text{cond}}^{\,\text{Si}}\rangle}}
\newcommand{\mfcondmg}{\ensuremath{\langle f_{\text{cond}}^{\,\text{Mg}}\rangle}}
\newcommand{\mfcondsim}{\ensuremath{\langle\widetilde{f}_{\text{cond}}^{\,\,\text{Si}}\rangle}}
\newcommand{\mfcondmgm}{\ensuremath{\langle\widetilde{f}_{\text{cond}}^{\,\,\text{Mg}}\rangle}}
\newcommand{\mdrhog}{\ensuremath{\langle\drhog\rangle}}
\newcommand{\mdrhogf}{\ensuremath{\left\langle\drhogf\right\rangle}}

\newcommand{\mdrad}{\ensuremath{\langle\radd\rangle}}
\newcommand{\mvdrinf}{\ensuremath{\langle\vdrinf\rangle}}
\newcommand{\mradd}[1]{\ensuremath{\langle a_{\text{#1}}\rangle}}

\newcommand{\mvDinf}{\ensuremath{\langle\vDinf\rangle}}

\newcommand{\Rs}{\ensuremath{R_{\star}}}
\newcommand{\Ms}{\ensuremath{M_\star}}
\newcommand{\Ls}{\ensuremath{L_\star}}
\newcommand{\Lsun}{\ensuremath{\text{L}_{\sun}}}
\newcommand{\Msun}{\ensuremath{\text{M}_{\sun}}}

\newcommand{\rfluc}{\ensuremath{\hat{r}}}
\newcommand{\pP}{\ensuremath{P}}

\newcommand{\rSaHoc}{Paper~III}

\newcommand{\rSaMa}{Paper~V}
\newcommand{\rMaSa}{MS21}

\newcommand{\rGS}{GS14}

\newcommand{\rHo}{H08}
\newcommand{\rBl}{B19}

\begin{document}

   \title{Three-component modelling of O-rich AGB star winds}

   \subtitle{I. Effects of drift using forsterite\thanks{All input parameter files, the used custom opacity data file, as well as all resulting model files are, together with the tools to read the files, openly available at \href{https://doi.org/10.5281/zenodo.8051249}{https://doi.org/10.5281/zenodo.8051249}.}}

   \author{C. Sandin\inst{1}
          \and
          L. Mattsson\inst{2}
          \and
          K.~L. Chubb\inst{3}
          \and
          M. Ergon\inst{4, 1}
          \and
          P.~M.\ Weilbacher\inst{5}
   }

   \institute{Department of Astronomy, AlbaNova University Center,
     Stockholm University, SE-10691 Stockholm, Sweden\\
     \email{christer.sandin@astro.su.se}
    \and
     Nordita, KTH Royal Institute of Technology and Stockholm University,
     Hannes Alfv{\'e}ns v{\"a}g 12, SE-10691 Stockholm, Sweden\\
     \and
     Centre for Exoplanet Science, University of St Andrews, North Haugh,
     St Andrews, KY16 9SS, United Kingdom\\
     \and
     The Oskar Klein Centre, AlbaNova, SE-10691 Stockholm, Sweden\\     
     \and
     Leibniz-Institut f{\"u}r Astrophysik Potsdam (AIP),
     An der Sternwarte 16, 14482 Potsdam, Germany\\
   }

   \date{Submitted 3 January 2023; accepted 13 June 2023}

   \abstract%
   {Stellar winds of cool and pulsating asymptotic giant branch (AGB) stars enrich the interstellar medium with large amounts of processed elements and various types of dust. %}
   We present the first study on the influence of gas-to-dust drift on ab initio simulations of stellar winds of M-type stars driven by radiation pressure on forsterite particles. %}
   Our study is based on our radiation hydrodynamic model code {\teh} that includes frequency-dependent radiative transfer, dust extinction based on Mie scattering, grain growth and ablation, gas-to-dust drift using one mean grain size, a piston that simulates stellar pulsations, and an accurate high spatial resolution numerical scheme. To enable this study, we calculated new gas opacities based on the {\exomol} database, and we extended the model code to handle the formation of minerals that may form in M-type stars. We determine the effects of drift by comparing drift models to our new and extant non-drift models. %}
   Three out of four new drift models show high drift velocities,
   $87$--$310\,\kms$. Our new drift model mass-loss rates are 1.7--13\, per cent of the corresponding values of our non-drift models, but compared to the results of two extant non-drift models that use the same stellar parameters, these same values are 0.33--1.5 per cent. Meanwhile, a comparison of other properties such as the expansion velocity and grain size show similar values. Our results, which are based on single-component forsterite particles, show that the inclusion of gas-to-drift is of fundamental importance in stellar wind models driven by such transparent grains. Assuming that the drift velocity is insignificant, properties such as the mass-loss rate may be off from more realistic values by a factor of  50 or more.}


   \maketitle
%
%________________________________________________________________

\section{Introduction}\label{sec:introduction}
Stellar winds dominate the final and decisive stages of evolution of low- to intermediate-mass stars when they ascend the asymptotic giant branch (AGB). The dynamic AGB stage involves increasing luminosities, low effective temperatures, and stellar pulsations. Dust formation begins at about two stellar radii where temperatures are low enough to prevent the newly formed grains from evaporating, and new dust grains absorb or scatter the radiation and in that way attain  momentum. The grains accelerate outwards and collide with particles in the gas that are dragged along as the particles drift through the same gas. Considering all the needed physics, it is a complex physical problem to simulate the resulting dust-driven wind where low expansion velocities are about $10\kms$ and high mass-loss rates vary from $10^{-8}$ up to, in extreme cases, $10^{-4}\,\mdotu$.

Depending on what element dominates, AGB stars are either oxygen-rich (M-type stars) or carbon-rich (C-type stars). The dichotomy is reflected in stellar wind models where dust formation in a carbon-rich chemistry is more simple where mostly amorphous carbon forms. Other types of dust and minerals do not appear to form in sufficient numbers to be influential.

Dust formation in an oxygen-rich chemistry is more complex. Spectra of circumstellar envelopes of M-type AGB stars show characteristic silicate features at 9.7 and $18\,\mu$m  \citep[see e.g.][]{WoNe:69,Lo:70,MoWaTi:02,Do:10,MoWaKe:10}. These features indicate that silicon-containing grains are a dominant component in M-type AGB stars. Crystalline silicate dust with features at 11, 23, 28, 33, and $69\,\mu$m is also seen \citep{BlVrWa.:14}, but the crystallinity does not appear to be correlated with mass-loss rates \citep{LiJiLiGa:17}. Various minerals form depending on the availability of elements that are part of the different minerals, including olivine, pyroxene, and iron \citep{GaSe:99}.

Metallic iron, moreover, appears to be a significant component in the cosmic dust budget owing to the large iron depletion seen in the interstellar medium \citep{MaDeCia:19}. These grains can probably form in AGB atmospheres and their scattering cross-sections are typically large, so if they form in sufficient number, they may contribute to the driving of the wind. \citet[hereafter {\rGS}]{GaSe:14} present a refined and in many ways complete (rates-based) approach on how to implement mineral formation in both carbon- and oxygen-rich chemistries.

\citet[hereafter {\rHo}]{Ho:08} presents the first working models (\darwin) of stellar winds in oxygen-rich chemistry. She finds that the dust scattering  cross-section of larger micron-sized iron-free silicate particles provide a high enough radiative pressure to drive a stellar wind. \citet{BlHo:12} and \citet{BlHoNo.:13} then argue, based mostly on  parametrised models of dust, that forsterite and enstatite are the most likely dust species that drive the stellar wind; they also present photometric properties of models that closely agree  with observations. \citet{BlHoArEr:15} present a larger set of radiation hydrodynamic models that include non-equilibrium dust formation. The authors conclude that they can calculate mass-loss rates as well as spectra; their visual and near-IR diagnostics agree with observations. \citet[hereafter {\rBl}]{BlLiHo.:19} present the most extensive set of calculated M-type stellar wind models available to date. Whilst the stellar wind models of {\rHo} up to {\rBl} show agreement with observations, they are based on some assumptions that we find interesting to explore in more detail. The authors emphasise that they calculate high mass-loss rates and photometric properties that agree well with  observations. They also point out that there are few free parameters in their radiation hydrodynamic models. In particular, the only such free parameter they mention is the seed particle abundance. The authors, moreover, appear to use sticking coefficients that are always set to unity (1) to form as much dust as possible, instead of using extant lower empirically based values. Additional assumptions include only  modelling one (or two) dust species at a time.

Physical arguments imply that the effects of drift are stronger in these winds than in carbon-rich environments \citep[hereafter {\rMaSa}]{MaSa:21}. Only one extant study addresses the effects of drift, whilst assuming very low drift velocities, lacking any evidence of higher values \citep{ToDeHuVe:22}. We think there is good reason to check the influence of drift on results more carefully. As we show here, drift velocities turn out to be dramatically higher in models of M-type stars than in models of C-type stars. Correspondingly, we also find dramatically lower mass-loss rates. With our physically and numerically extended models, we are  unable to reproduce the higher  mass-loss rates of {\darwin} on which  the authors base their results of good agreement with observations. Whilst more reliable observations of mass loss show higher mass-loss rates, it is clear that something important is missing in the picture of understanding the formation of stellar winds in M-type stars.

Extant ab initio stellar wind models that include drift are all based on a carbon-rich chemistry. \citet[hereafter {\rSaMa}]{SaMa:20} include frequency-dependent radiative transfer and opacity tables of both the gas and the dust, and calculate models at high spatial resolution. The results indicate important differences between drift models and position-coupled (PC, i.e. non-drift) models. Mass-loss rates, expansion velocities, and yields of dust are affected. An additional example of a carbon-rich model where drift is found to be an important component to understand the observations is presented in a study of grain alignment about \object{IRC$+10^\circ216$} \citep{AnLoMe.:22}; this object shows a very high mass-loss rate (of 2--8$\times10^{-5}\,\Msun\,\text{yr}^{-1}$), where our model nevertheless shows a drift velocity that is twice as high as the expansion velocity.

We use our simulation code {\teh} of {\rSaMa} and extend it with the rates-based description of dust formation in an oxygen-rich chemistry. Specifically,   here we focus on a wind where only forsterite is formed. To enable this study, we calculated new gas opacity tables for solar metallicities based on the {\exomol} database~\citep{2020JQSRT.25507228T}, and also added free-free and bound-free opacities calculated using the \textsc{jekyll} code \citep{ErFrJe.:18,ErFr:22}. We are thereby able, for the first time,  to study time-dependent models using high spatial resolution in simulations that use an oxygen-rich chemistry that includes drift.

We first make semi-analytical predictions of the drift velocity in Sect.~\ref{sec:semianal} to see what we can deduce based on simple physical arguments. Thereafter, we describe the physics features of our physically enhanced models in Sect.~\ref{sec:model}. Presentations of the modelling procedure and results follow in Sect.~\ref{sec:modres}. We discuss the influence of drift on our results in Sect.~\ref{sec:discussion}, and close the paper with our conclusions in Sect.~\ref{sec:conclusions}.

\section{Semi-analytic predictions of drift}\label{sec:semianal}
Before we engage in numerical and physical details of our updated version of {\teh}, we look at a simplified treatment of the oxygen-rich stellar-wind formation problem to estimate how high associated drift velocities could be. We address the balancing forces that give rise to the wind in Sect.~\ref{sec:analforce}. Thereafter we look again at the concept of complete momentum coupling in Sect.~\ref{sec:analcmc}, and conclude this analysis in Sect.~\ref{sec:analinterp}.

\subsection{Balancing dust extinction and radiation pressure}\label{sec:analforce}
To estimate the drift velocity for a given set of stellar parameters, we need to estimate the radiation pressure on the dust component. Thus, we need to know the photon-to-dust grain momentum transfer efficiency. Absorption and scattering of photons by dust grains are modelled with the effective cross-sections $\sigma_{\text{abs},\,\nu}$ and $\sigma_{\text{sca},\,\nu}$, where $\nu$ is the frequency. The efficiency of absorption and scattering, or the combination of the two (extinction), is usually defined relative to the geometric cross-section $\sigma$. For spherical grains, $\sigma=\pi a^2$, where $a$ is the grain radius. The absorption efficiency ${\Qnuabs}=\sigma_{\text{abs},\,\nu}/\sigma$ is related to the extinction and scattering efficiencies ${\Qnuext}=\sigma_{\text{ext},\,\nu}/\sigma$ and ${\Qnusca}=\sigma_{\text{sca},\,\nu}/\sigma$, as
\begin{equation}
  \Qnuabs=\Qnuext-\Qnusca.
\end{equation}
To calculate a correct radiation pressure, it is necessary to use the absorption efficiency (named  {\Qnuabspr} in {\rSaMa}),
\begin{equation}
  Q_{\text{rp},\,\nu}=\Qnuext-g_{\text{sca},\,\nu}\Qnusca = \Qnuabs - (g_{\text{sca},\,\nu} - 1)\,\Qnusca,
\end{equation}
where $g_{\text{sca},\,\nu}=\cost_{\nu}$ is the average scattering angle. Mie theory \citep{BoHu:83} provides these efficiencies and the average scattering angle.

\begin{figure*}
    \centering
    \sidecaption
    %\resizebox{\hsize}{!}{
    \includegraphics[trim=1.0cm 0.5cm 1.0cm 2.5cm, clip=true, width=12.0cm]{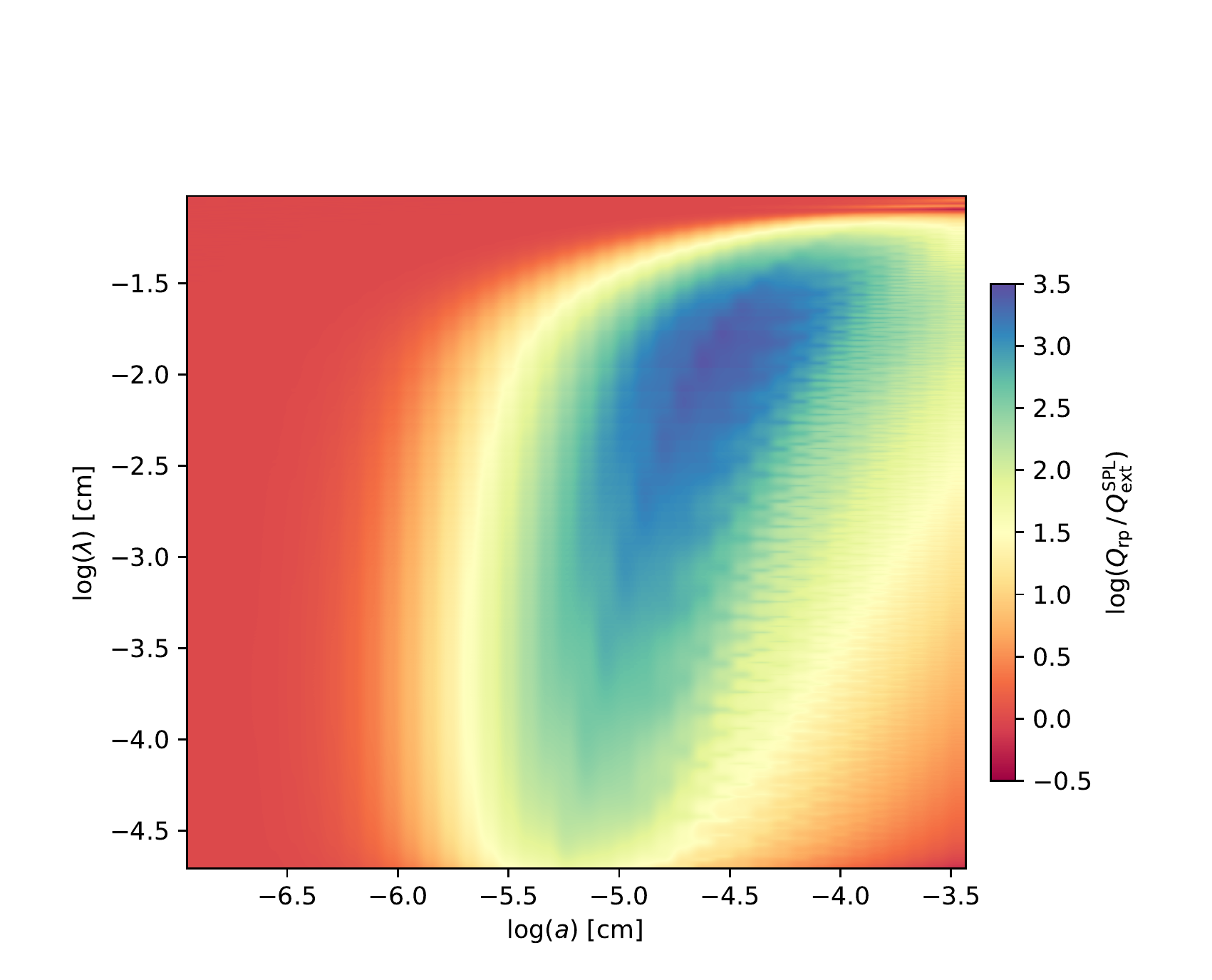}%}
    \caption{Ratio of the radiation pressure efficiency $Q_{\text{rp},\,\nu}$ to the SPL extinction efficiency vs   grain radius and wavelength, assuming spherical forsterite grains.}
    \label{fig:qrp}
\end{figure*}

In many extant works on AGB winds \citep[e.g.][]{SaHo:03,MaWaHo.:08,MaWaHo:10}, the radiation pressure is calculated assuming that dust grains are small compared to the wavelength of the incident radiation. This is called the small-particle limit (SPL) approximation, and it leads to the simplification $\Qnuext=Q_{\nu}^{\prime}/a$, where $Q_{\nu}^{\prime}$ is a function of only the frequency \citep{Wi:72}. Figure~\ref{fig:qrp} shows how $Q_{\text{rp},\,\nu}$, as computed based on Mie theory, compares to the corresponding SPL value of $\Qnuext$, using the optical constants of forsterite of \citet{JaDoMu.:03}. An important feature is the blue peacock feather-like region where the Mie theory-based radiation force on grains of radius $a$ in an optically thin atmosphere is (where the Eddington flux $H_\nu(r)\approx 0.25(R_\star/r)^2$)
\begin{equation}
    \fradd=\frac{\pi}{c}\left(\frac{R_{\star}}{r}\right)^2\,n_{\text{d}}\left(a\right)\int_{0}^{\infty}a^{2}\, Q_{\text{rp},\,\nu}\left(a\right)\,B_\nu(\teff)\,\text{d}\nu.
    \label{eq:frad}
\end{equation}
Here $R_\star$ is the stellar radius, $n_{\text d}$ the grain number density, and $B_\nu$ the Planck function. Grain sizes in the region of relevant values ($0.1\la a\la0.5\,\mu$m) result in a radiation pressure about 300--30 times lower in the spectral region near the typical flux peak of M-type AGB stars ($\lambda\simeq1\,\mu$m) than when the SPL is assumed.

\subsection{Momentum coupling and equilibrium drift}\label{sec:analcmc}
In {\rSaMa} complete momentum coupling (CMC) is defined as the case of force balance between radiation on the one hand and drag and gravity on the other, although the amount of momentum lost owing to the gravitational potential is negligible. Equating the radiation and drag forces is also a common definition of CMC. 

Drift is a non-linear dynamic phenomenon, but simulations (in particular those described in \rSaMa) show that an equilibrium tends to develop in most cases. Equilibrium drift can be defined as the situation where the Lagrangian derivatives of gas and dust velocity are equal (i.e. $\text{d}v/\text{d}t=\text{d}u/\text{d}t$). This, in turn, means that the equilibrium drift velocity $\vDeq$ is constant with respect to time and that $\vDeq$ is governed by a simple algebraic equation instead of a hard-to-solve partial differential equation.

Assuming  equilibrium drift and CMC defined as above, the drag force is $\fdrag=\fradd-\fgravd$. We define the dimensionless variable as
\begin{eqnarray*}
S_{\text{D}}=\frac{\vDeq}{\vzeta},\quad\text{and}\quad\vzeta=\sqrt{\zeta\Tg}=\sqrt{\frac{128\kB}{9\pi\mu m_{\text{H}}}\Tg},
\end{eqnarray*}
where $\vzeta$ is a modified thermal velocity, $\kB$ is the Boltzmann constant, $\mu$ the mean molecular weight, $m_{\text{H}}$ the mass of a hydrogen atom, and {\Tg} the gas temperature. We then have
\begin{equation}
S_{\text{D}}^2=-\frac{1}{2}+\left\{\frac{1}{4}+\left(\frac{\displaystyle f_{\text{rad, d}}-f_{\text{grav, d}}} {f_{\zeta}}\right)^2\right\}^{\frac{1}{2}},\label{eq:sdi}
\end{equation}
where $f_{\text{grav,d}}\approx\rhod\,GM_\star r^{-2}$ is the point-mass approximation for the gravitational force, {\rhod} the dust density, $G$ the gravitational constant, and $f_{\zeta}=\pi a^2\,n_{\text{d}}\,\rhog\,\vzeta^2$ can be seen as a thermal coupling force between gas and dust (where {\rhog} is the gas density). For a given mass-loss rate $\dot{M}$, wind expansion velocity {\uinf}, luminosity $L_\star$, effective temperature $\teff$, and stellar mass $M_\star$, we can now estimate the equilibrium drift velocity $\vDeq$ using the above equation in combination with the condition for mass conservation $\dot{M} = 4\pi\,r^2\,\rhog\uinf$ and $L_\star = 4\pi\,\sigma_{\text{SB}}\,R_\star^2\,\teff^4$, where $\sigma_{\text{SB}}$ is the Stefan-Boltzmann constant.

\begin{figure*}
    \centering
    \resizebox{\hsize}{!}{
    \includegraphics[trim=0.9cm 0.6cm 6.72cm 1.45cm, clip=true, height=11.8cm]{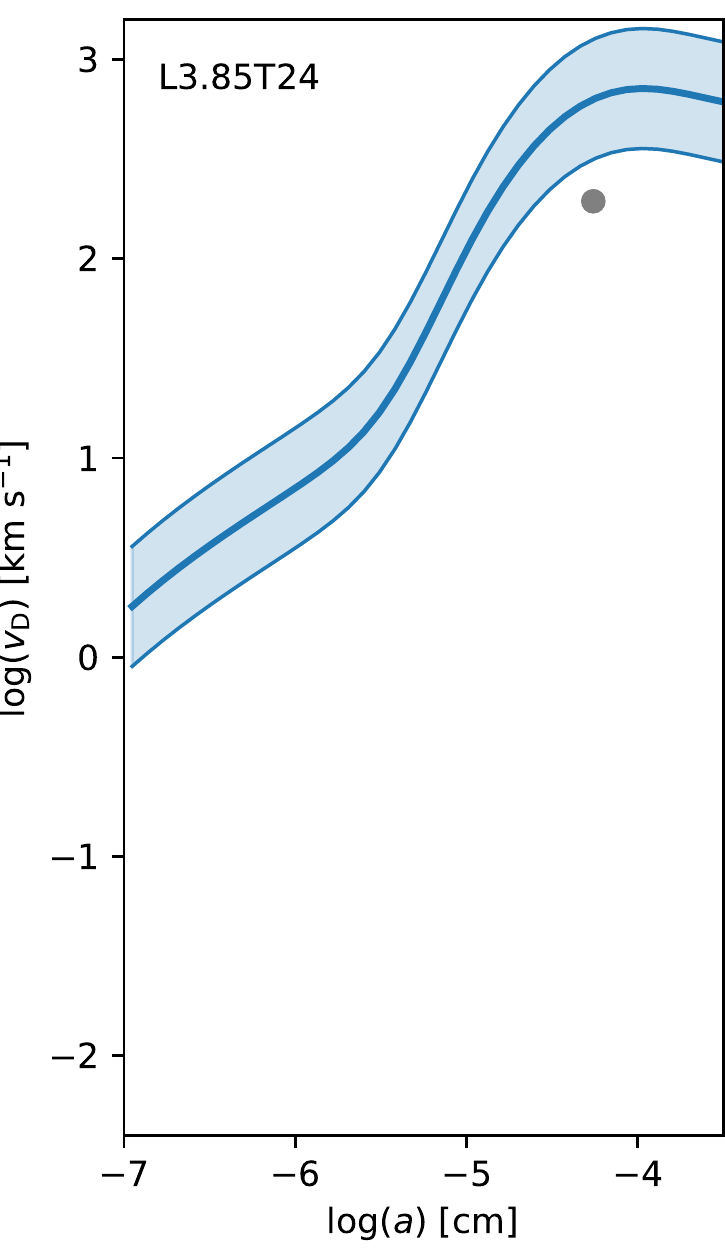}
    \includegraphics[trim=2.15cm 0.6cm 6.725cm 1.45cm, clip=true, height=11.8cm]{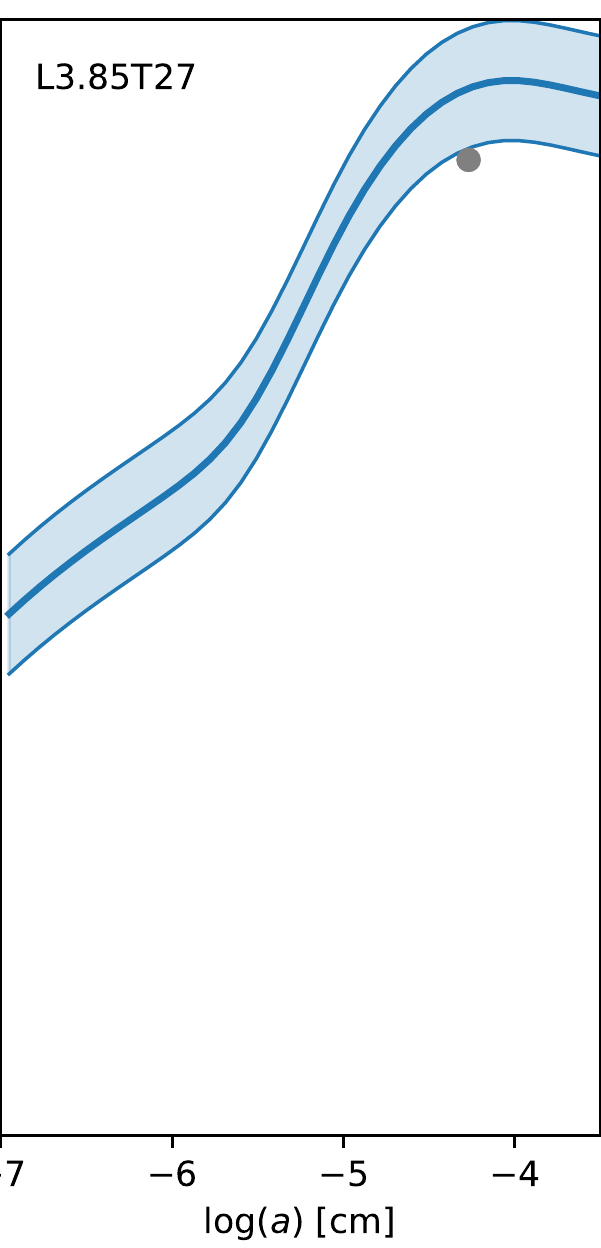}
    \includegraphics[trim=2.15cm 0.6cm 6.72cm 1.45cm, clip=true, height=11.8cm]{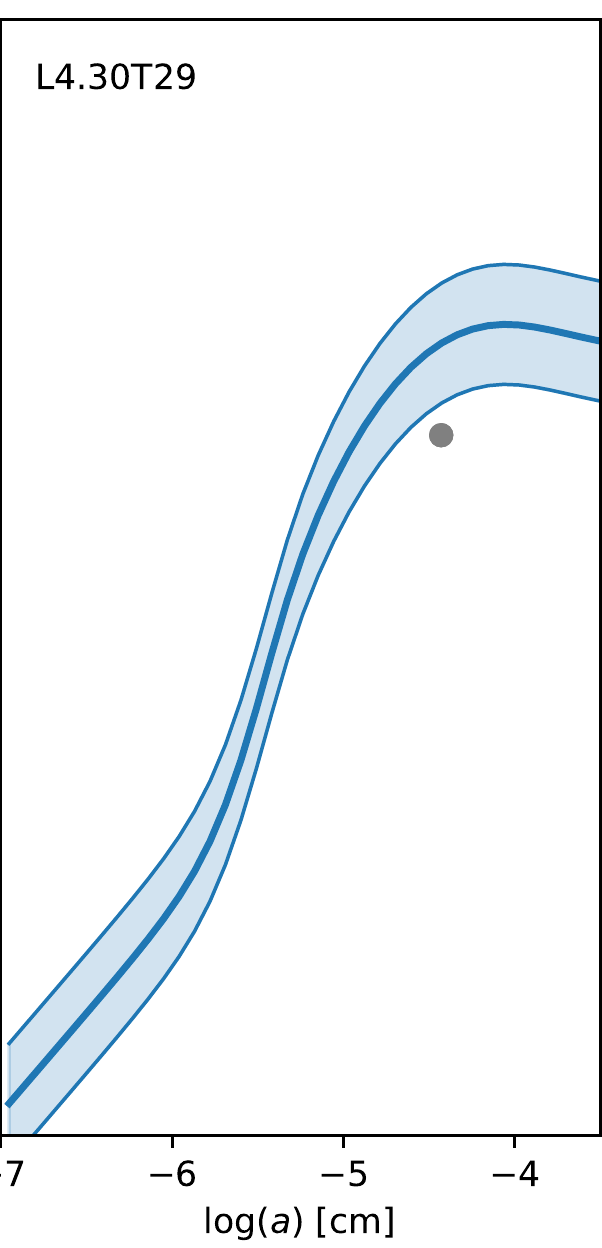}
    \includegraphics[trim=2.15cm 0.6cm 6.72cm 1.45cm, clip=true, height=11.8cm]{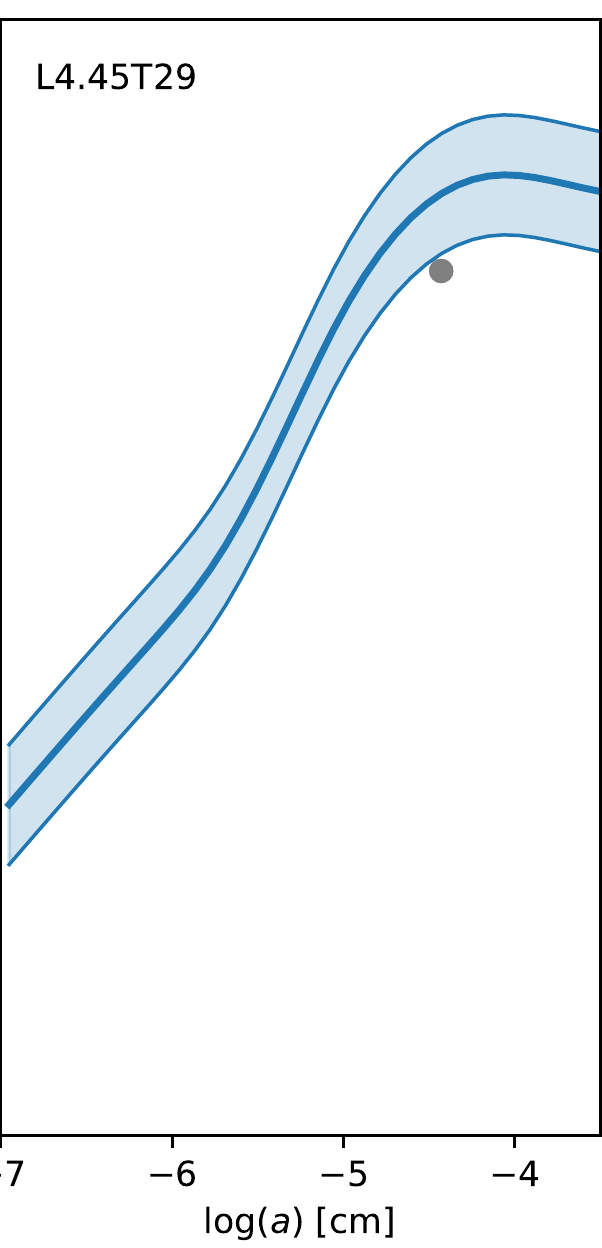}
    }
    \caption{
      Equilibrium drift velocity $\vDeq$ vs grain radius, assuming CMC. The four panels show different parameter configurations considered in the present study; all panels use the same ordinate range. The shaded regions show factor of 2 variations in the semi-analytical results based on equilibrium theory (solid lines). The bullets indicate the locations of the two numerical simulations with drift presented in Sect.~\ref{sec:results}.
    \label{fig:vDeq}}
\end{figure*}

\subsubsection{Predicted drift velocities}
In Fig. \ref{fig:vDeq} we show the expected equilibrium drift velocity $\vDeq$ versus grain radius $a$ for the $M_{\star}$, $\log(L_{\star})$, $\teff$, and mass-loss rates of our detailed simulations (presented in Sect.~\ref{sec:results}). The stellar parameters were chosen to represent recent stellar evolution models \citep[in particular those of][]{MaDeKa:23}, but also observations \citep[see e.g.][and references therein]{UtMcDetal:19}. The predictions from equilibrium theory are in acceptable agreement with our detailed modelling, overpredicting $\vD$ by just a little more than a factor of two. Thus, we conclude there is a solid case for high drift velocities and that the maximum drift velocity of each curve occurs for grain radii that are quite typical for M-type AGB stars according to extant models (\rBl).

It is clear that a wind driven by radiation pressure on forsterite grains leads to high drift velocities, even in case of massive outflows. It has been argued that drift is negligible in winds associated with very high mass-loss rates \citep{HoOl:18}, which seems to be the case for carbon stars (see Fig.~3a in {\rSaMa}); however, our more recent model of the stellar wind in IRC$+10^\circ216$ says otherwise \citep[see Sect.~4.3 and Fig.~8 in][]{AnLoMe.:22}. Given that M-type AGB stars have wind speeds of $u_{\infty}\sim 10\,\kms$ \citep[as  revealed by the radio observations of][]{OlGoKe:02,GoOlKe:03} and typical grain sizes $a\simeq0.5\,\mu$m, the drift factor $\FDi=1+\vD/\uinf>2$ also for intense outflows. A drift factor $\FDi>2$ corresponds to a situation where the drift velocity $\vD$ exceeds the gas expansion velocity $\uinf$ and the dust mass-loss rate increases by as much when compared to the gas  mass-loss rate (see Eq.~(32) in {\rSaMa)}. We tentatively conclude that including drift in the modelling of M-type AGB stars is of  fundamental importance, as PC models do not provide a correct result even in the high-mass-loss limit. As we show here, this conclusion is confirmed by our detailed modelling, described below. 

\subsection{Physical interpretation of the PC assumption}\label{sec:analinterp}
The PC assumption is incompatible with the idea that AGB winds form by friction between radiatively accelerated dust grains and gas particles. {\rMaSa} argue that there is no realistic physical limiting case which leads to PC. Whilst this must still be true, we    discuss here the (unrealistic) limit where PC is formally true. 

Considering Eq.~(\ref{eq:sdi}), we note that $(\fradd-\fgravd)/{f_{\zeta}} \ll 1$ implies $S_{\text{D}}=0$ (i.e. no drift). If we ignore the case $\fradd=\fgravd$, this limit requires that $f_\zeta$ is very large and, in particular, much larger than the net radiation force $\tilde{f}_{\text{rad,d}}=\fradd-\fgravd$. For this to occur in the case given in the previous section (with {\fradd} obtained from Eq.~(\ref{eq:frad})), the modified thermal velocity {\vzeta} has to be of the order of $100\,\kms$ unless {\rhog} is several orders of magnitude higher than expected in a realistic wind. Such a high {\vzeta} corresponds to gas temperatures of the order of $10^5\,$K, which is entirely unrealistic. It is fair to say that the physical interpretation (or consequence) of the assumption of PC is absurd. The semi-analytic predictions of drift velocity presented here provide a solid theoretically founded reason for pursuing detailed modelling of winds of M-type AGB stars with gas and dust treated as dynamically decoupled phases.

\section{Model features and improvements {\bf of T-800}}\label{sec:model}
The model features of our radiation hydrodynamic model code {\teh} is described in {\rSaMa}. As in the case of a C-rich chemistry, we model three components in the O-rich chemistry described here: the gas, the radiation field, and a dust component consisting of forsterite (Fo) mineral grains.

In comparison to the moments method used to describe dust formation in the C-rich chemistry, we replace the four dust moment equations ($K_0$--$K_3$) with one rate equation for the formation of each mineral $k$, and the carbon number density equation (Eq.~(5) in {\rSaMa}) with corresponding equations for each affected tracer element $j$. All the physics of mineral formation in oxygen-rich chemistry that we require is developed and described in \citet{GaSe:99} and {\rGS}. The adjusted equations are
\begin{eqnarray}
\frac{\partial}{\partial t}\nds{k}N_{k}+\nabla\cdot\left(\nds{k}N_{k}v_{k}\right)&=&q_{k},\\
\frac{\partial}{\partial t}n_{j}+\nabla\cdot\left(n_{j}u\right)&=&-\nu_{k}^{j}q_{k},\label{eq:ndenj}
\end{eqnarray}
where $t$ is the time, $\nds{k}$ the seed particle density, $N_{k}$ the number of monomers, $v_{k}$ the mean dust particle velocity, $q_{k}$ the sum of the source and sink terms owing to grain formation, $n_{j}$ the tracer element atom number density, $\nu_{k}^{j}$ the number of tracer element atoms per monomer, and $u$ the gas velocity. In this approach there is no description of nucleation. Instead, seed particles of each individual species are assumed to exist everywhere and
\begin{eqnarray}
\nds{k}=\rhog\frac{\epsilon_{k}}{m_{\text{u}}\mu},
\end{eqnarray}
where $\epsilon_{k}$ the seed particle abundance, $m_{\text{u}}$ the atomic mass constant, and $\mu=1.26$.

\subsection{Mineral rate equation}
The rate equations track the number of monomers $N_{k}$ throughout the model domain. The rate equation source term owing to grain growth, evaporation, and destruction is
\begin{eqnarray}
  q_{k}=\nds{k}\left[4\pi a_{k}^2\left(J_{\text{gr},k} - J_{\text{ev},k}\right)
    - \frac{1}{\tau_{\text{sp,n}}^{k}}\right],\label{eq:rate}
\end{eqnarray}
where $a_{k}$ is the particle radius, $J_{\text{gr}}$ ($J_{\text{ev}}$) the growth (evaporation) rate, and $1/\tau_{\text{sp,n}}$ the rate of non-thermal sputtering. The grain radius is described using the grain volume $V_k$ and the monomer volume $V_1$,
\begin{eqnarray}
a_{k}=\left(\frac{3V_{k}}{4\pi}\right)^{\frac{1}{3}},\quad V_{k}=N_{k}V_{1,k},\quad V_{1,k}=\frac{\mathcal{A}_{k}m_{\text{u}}}{\rhoint{k}},
\end{eqnarray}
where $\mathcal{A}_{k}$ is the molecular weight and $\rhoint{k}$ the mineral intrinsic density.

The term that describes grain growth and evaporation is written as (\rGS, Eqs.~(12.101), (12.102), and (12.108))
\begin{eqnarray}
  J_{\text{gr}, k} - J_{\text{ev}, k}=\xi_{k}\frac{p_{j}}{\sqrt{2\pi m_{j}\kB\Tg}}\left(\phi_{k}-\frac{1}{a^{\text{c}}_{k}}\sqrt{\frac{\Tg}{\Td}}\right),
\end{eqnarray}
where $\xi_{k}$ is the the drift-velocity-dependent sticking coefficient, and  $p_{j}$ and $m_{j}$ are the partial pressure and mass of the rates-determining component, respectively (see Sect.~\ref{sec:grevap}). Moreover, $\phi_{k}$ is the drift correction factor, $a^{\text{c}}_{k}$ the reaction activity, and {\Td} the dust temperature.

The sticking coefficient\footnote{In {\rMaSa}, $\xi$ is used to denote the grain-growth velocity, which is a different, although not unrelated, quantity.} $\xi_{k}$ is assumed to decrease when the drift velocity becomes high in relation to the binding energy $E_{\text{b},k}$~(Eq.~(14) in \citealt{KrSe:97} and Eq.~(13) in \citealt{SaHo:04}, hereafter {\rSaHoc}):
\begin{eqnarray}
 \xi_{k}=\xi^{(k)}\exp\left[-\left(\frac{\mathcal{A}_{k}m_{\text{u}}\tilde{w}^2_{k}}{8E_{\text{b},k}}\right)^3\right],\label{eq:xik}
\end{eqnarray}
where the velocity of dust grains relative to gas particles $\tilde{w}_{k}$ is (Eqs. (11) and (12) in {\rSaHoc})
\begin{eqnarray}
\tilde{w}_{k}=\left(\frac{8\kB\Tg}{16\pi\mathcal{A}_{k}m_{\text{u}}}+\frac{v^2_{\text{D},k}}{16}\right)^{\frac{1}{2}}.
\end{eqnarray}
Here, $v_{\text{D},k}=v_{k}-u$ is the drift velocity. Furthermore, the drift correction factor $\phi_{k}$ is (Eq.~(12.19) in \rGS)
\begin{eqnarray}
\phi_{k}=\left(1+\frac{\pi\mathcal{A}_{k}m_{\text{u}}}{8\kB}\frac{v^2_{\text{D},k}}{\Tg}\right)^{\frac{1}{2}}.
\end{eqnarray}
We use the same expression for non-thermal sputtering ($1/\tau_{\text{sp,n}}^{k}$) as we do in {\rSaHoc}. Although, here we account for collisions with H$_2$ molecules in addition to H and He atoms.

\subsection{Growth and evaporation of forsterite}\label{sec:grevap}
We use two tracer elements: silicon and magnesium. There are in this case 11 equations, instead of 13 equations when using the moments approach and a carbon-rich chemistry. Forsterite grain growth takes place through collisions of seed grains and extant grains with either SiO molecules or Mg atoms; when the addition of SiO (Mg) is the rate determining reaction step, $p_{j}=p_{\text{SiO}}$ and $m_{j}=m_{\text{SiO}}$ ($p_{j}=p_{\text{Mg}}$ and $m_{j}=m_{\text{Mg}}$).

The basic chemical reaction for forsterite formation, as well as its evaporation through chemical sputtering, is
\begin{eqnarray}
  2\text{Mg} + \text{SiO} + 3\text{H}_2\text{O} \leftrightarrow \text{Mg}_2\text{SiO}_4(s) + 3\text{H}_2
\end{eqnarray}
and the (chemical sputtering) reaction activity $a^{\text{c}}_{\text{Fo}}$ is (see Eqs.~(12.60), (12.103), and (12.104) in {\rGS})
\begin{eqnarray}
  \frac{1}{a^{\text{c}}_{\text{Fo}}}=\frac{p_{\text{H}_2}^3}{p_{\text{SiO}}p_{\text{Mg}}^2p_{\text{H}_2\text{O}}^3}\frac{\Kp{\text{SiO}}\Kps{3}{\text{H}_2\text{O}}}{\Kp{\text{Mg}_2\text{SiO}_4}\Kps{3}{\text{H}_2}},
\end{eqnarray}
where the four equilibrium constants $K_{\text{p}}$ are calculated at the dust temperature {\Td}.

\subsection{Partial pressures of atoms and molecules}\label{sec:ppress}
The number densities of the molecules in the gas phase that are part of the grain formation as well as the activities that determine when dust grains form are calculated in an equilibrium chemistry of molecules with hydrogen, oxygen, carbon, nitrogen, aluminium, silicon, and sulphur, following the approach of {\rGS} (chapter 10.3). The considered atoms and molecules are H, H$_2$, O, OH, H$_2$O, CO, CO$_2$, CH$_4$, N, N$_2$, NH$_3$, HCN, Al, AlO, AlS, AlOH, AlO$_2$H, Al$_2$O, Al$_2$O$_2$, Si, SiO, SiO$_2$, S, SO, HS, H$_2$S, SiS, and S$_2$. Magnesium is assumed to be present as free atoms.

All number densities and activities are calculated for the temperature range $100\le\Tg\le10\,000\,$K. We use the equilibrium constants $K_{\text{p}}$, which  are often referred to as dissociation constants,  of \citet{ShHu:90}, \citet{BaCo:16}, {\rGS} (see their Table~A.5), and NIST JANAF.\footnote{The equilibrium-constant data of NIST/JANAF can be retrieved from \href{https://janaf.nist.gov/}{https://janaf.nist.gov/}.}

\section{Modelling procedure and results}\label{sec:modres}
We first briefly describe  our modelling procedure in Sect.~\ref{sec:modelling} and then describe the physics  set-up and choice of model parameter sets in Sect.~\ref{sec:modelpar}. We present our results in Sect.~\ref{sec:results}.

\subsection{Modelling procedure}\label{sec:modelling}
We follow the modelling procedure described in Sect.~3.1 in {\rSaMa}. Due to the low outflow velocity of the wind ($\uinf\la10\,\kms$), we set the outer boundary here at $\rextf=20\Rs$. We use $N_{\text{d}}=840$ grid points, which very nearly corresponds to the grid point arrangement we achieve when using $N_{\text{d}}=1024$ and $\rextf=40\Rs$.
It appears to be sufficient to evolve the wind models for a time interval of about $100\,P$ (stellar pulsation periods) as the wind structures reach a state of equilibrium before that.

\begin{table}
\caption{References of the atom and molecule datasets used, and number of energy levels accounted for in each entry.}\label{tab:linedata}
\begin{tabular}{lrll}\hline\hline\\[-1.5ex]
molecule & dataset & References & Energy levels\\[1.0ex]\hline\\[-1.8ex]
         C &      Kurucz & 1          & 999\\
     C$_2$ &     8states & 2, 3       & 44\,189\\
C$_2$H$_2$ &       aCeTY & 4          & 5\,160\,803\\
        CH &     MoLLIST & 5, 6       & 2526\\
    CH$_4$ &    YT34to10 & 7, 8       & 8\,194\,057\\
        CN &   Trihybrid & 9, 10, 11  & 7703\\
        CO &      Li2015 & 12, 13     & 6383\\
    CO$_2$ &    UCL-4000 & 14         & 3\,562\,798\\
        CS &         JnK & 15         & 11497\\
       CrH &     MoLLIST & 6          & 1646\\
       FeH &     MoLLIST & 16, 6      & 3564\\
     H$_2$ &      RACPPK & 17         & 302\\
    H$_2$O &   POKAZATEL & 18         & 810\,269\\
    H$_2$S &        AYT2 & 19         & 220\,618\\
       HCl &      HITRAN & 20         & 335\\
       HCN &      Harris & 21, 22     & 168\,110\\
        HF & Coxon-Hajig & 23, 24, 13 & 684\\
       LaO &         BDL & 25         & 38\,208\\
       MgH &         XAB & 26         & 1303\\
         N &      Kurucz & 1          & 283\\
     N$_2$ &      WCCRMT & 27, 28, 29 & 40\,380\\
    NH$_3$ &      CoYuTe & 30, 31     & 5\,095\,730\\
         O &      Kurucz & 1          & 201\\
        OH &     MoLLIST & 32, 33, 6  & 1878\\
    SO$_2$ &     ExoAmes & 34         & 3\,270\,270\\
       SiO &   SiOUVenIR & 35         & 174\,250\\
       SiS &        UCTY & 36         & 10\,104\\
       TiH &     MoLLIST & 37         & 5788\\
       TiO &        Toto & 38         & 301\,370\\ % ?? ... was: 236\,308\\
        VO &       VOMYT & 39         & 638\,958\\
        YO &        SSYT & 40         & 79\,440\\
       ZrO &          SB & 41         & 3005\\[1.0ex]\hline
\end{tabular}
\tablebib{(1)~\citet{KURonline}; (2)~\citet{2018MNRAS.480.3397Y}; (3)~\citet{2020MNRAS.497.1081M}; (4)~\citet{2020MNRAS.493.1531C}; (5)~\citet{2014A&A...571A..47M}; (6)~\citet{2020JQSRT.24006687B}; (7)~\citet{2014MNRAS.440.1649Y}; (8)~\citet{2017A&A...605A..95Y}; (9)~\citet{2014ApJS..210...23B}; (10)~\citet{2020MNRAS.499...25S}; (11)~\citet{2021MNRAS.505.4383S}; (12)~\citet{2015ApJS..216...15L}; (13)~\citet{2021JChPh.155u4303S}; (14)~\citet{2020MNRAS.496.5282Y}; (15)~\citet{2015MNRAS.454.1931P}; (16)~\citet{2003ApJ...594..651D}; (17)~\citet{2019A&A...630A..58R}; (18)~\citet{2018MNRAS.480.2597P}; (19)~\citet{2016MNRAS.460.4063A}; (20)~\citet{2017JQSRT.203....3G}; (21)~\citet{2006MNRAS.367..400H}; (22)~\citet{2014MNRAS.437.1828B}; (23)~\citet{2013JQSRT.121...78L}; (24)~\citet{2015JQSRT.151..133C}; (25)~\citet{2022ApJ...933...99B}; (26)~\citet{2022MNRAS.511.5448O}; (27)~\citet{2018JQSRT.219..127W}; (28)~\citet{2017JQSRT.186..221W}; (29)~\citet{1969JChPh..51..689S}; (30)~\citet{2015JQSRT.161..117A}; (31)~\citet{2019MNRAS.490.4638C}; (32)~\citet{2016JQSRT.168..142B}; (33)~\citet{2018JQSRT.217..416Y}; (34)~\citet{2016MNRAS.459.3890U}; (35)~\citet{2022MNRAS.510..903Y}; (36)~\citet{2018MNRAS.477.1520U}; (37)~\citet{2005ApJ...624..988B}; (38)~\citet{2019MNRAS.488.2836M}; (39)~\citet{2016MNRAS.463..771M}; (40)~\citet{2019PCCP...2122794S}; (41)~\citet{2021ApJ...923..234S}}
\end{table}

\subsection{Physics set-up and selection of model parameters}\label{sec:modelpar}
We introduce effects of gas-to-dust drift using one mean dust velocity. We compare the new drift models to PC (non-drift) models that are in all other ways equivalent to the drift models.

We used the  solar abundances of \citet{AnGr:89}, with the values for C and O of \citet{GrSa:94}. Similarly to  {\rBl}, we set the pulsation period ($P$) using the $P$--$L_*$ relation of \citet{WhMeFe.:09}; this period-luminosity relation is based on observations of C-rich Mira stars in the Local Group Fornax galaxy. \citet{WhFeLe:08} present a relation of similar properties that would be more suitable to use with O-rich Mira stars (P.~Whitelock, priv. comm.). An accurately determined distance to  lower-metallicity stars in a Local Group galaxy allows a more accurate determination of the luminosity than for higher-metallicity stars in the Galaxy where distances are less well determined. It may be that there are differences between period-luminosity relations at different metallicities. \citet{Sa:23} argues that he finds such differences. We do not explore any metallicity dependence of the period-luminosity relation here considering our focus on effects of drift, but note that {\rBl} makes a test where the period is varied by about 10\% whereby the resulting mass-loss rates change by 20\%.

To correct for too small bolometric variations \citep{GaHoJoHr:04}, {\rBl} (see their Sect.~2.2) introduce a free parameter $f_{\text{L}}$ that allows larger variations of the luminosity at the inner boundary. We added the option to {\teh} to use a freely chosen value on $f_{\text{L}}$, and here we use $f_{\text{L}}=2$, although we note that for our purposes in this paper the results of models using either approach are indistinguishable.

Next, we describe our approach for calculating gas opacities in Sect.~\ref{sec:kappagas}, dust properties in Sect.~\ref{sec:propdust}, and our selection of model parameters in Sect.~\ref{sec:selmodel}.

\subsubsection{Gas opacities}\label{sec:kappagas}
In {\rSaMa} we used the tabulated gas opacities $\kappa_{\nu}\left(\rhog,\Tg\right)$ that were created for carbon-rich chemistries with the \textsc{coma} code \citep{BAr:00,ArGiNo.:09} for 319 wavenumbers in the interval $400\le\tilde{\nu}\le39\,480\,\text{cm}^{-1}$, 50 temperatures in the interval $1000\le\Tg\le10\,000\,$K, and 24 densities in the interval $-18\le\log_{10}\rhog\le-6\,\text{g}\,\text{cm}^{-3}$.

Here we calculated new bound-bound gas opacities based on data of the {\exomol} project \citep{2020JQSRT.25507228T}.\footnote{\href{https://www.exomol.com/}{https://www.exomol.com/}.} The calculations make use of data for the following 30 atoms and molecules: C, C$_2$, C$_2$H$_2$, CH, CH$_4$, CN, CO, CO$_2$, CS, CrH, FeH, H$_2$, H$_2$O, H$_2$S, HCl, HCN, HF, LaO, MgH, N, N$_2$, NH$_3$, O, OH, SO$_2$, SiO, SiS, TiH, TiO, VO, YO, and ZrO (see Table~\ref{tab:linedata}). The number of energy levels for each  dataset is specified here; however, the corresponding number of transitions or lines is typically at least an order of magnitude higher. For example, the {\exomol} aCeTY C$_2$H$_2$ line list has around 5.2 million energy levels and 4.3 billion transitions~\citep{2021A&A...646A..21C}. We used \textsc{exocross} \citep{YuAlTe:18}\footnote{\href{https://github.com/ExoMol/ExoCross}{https://github.com/ExoMol/ExoCross}.} to calculate  cross-sections $\sigma_{l}$ for each atom and molecule $l$ at 102\,750 wavenumbers in the interval $100\le\tilde{\nu}\le41\,200\,\text{cm}^{-1}$, 105 temperatures in the interval $100\le\Tg\le10\,000\,\text{K}$, and 24 gas densities in the interval $-18\le\log_{10}\rhog\le-6\,\text{g}\,\text{cm}^{-3}$. The cross-sections are resampled to a coarse grid of a pre-defined set of wavenumbers, where the resulting cross-section is the average of the ten nearest cross-sections on the finer grid. Currently, we are using 384 wavenumbers; this is an even multiple of the number of cores available on each node ($2\times64$) on the  high-performance cluster we used.
Individual cross-sections are thereafter converted to bound-bound opacities by multiplying by the corresponding partial pressure $p_{l}$ as
\begin{eqnarray}
\kappa_{\text{bb},\,l}=\frac{p_{l}}{\kB\Tg}\frac{\sigma_{l}}{\rhog}\,\left[\text{cm}^2\text{g}^{-1}\right].
\end{eqnarray}
We calculated partial pressures for the 27 molecules listed above as well as the three individual atoms using the same approach as in Sect.~\ref{sec:ppress}.

Bound-bound opacities become low at higher temperatures ($\Tg\ga2000\,$K), where instead bound-free opacities $\kappa_{\text{bf}}$ and free-free opacities $\kappa_{\text{ff}}$ dominate. We calculated these opacities using the \textsc{jekyll} code \citep{ErFrJe.:18,ErFr:22} (see Appendix~\ref{app:jekyll} for more information).

Finally, bound-bound and bound-free opacities of individual atoms, molecules, and ions as well as free-free opacities of ions are summed   to provide a total abundance-dependent gas opacity for each pair of gas density and gas temperature: 
\begin{eqnarray*}
\kappa_{\nu}\left(\rhog,\Tg\right)=\sum_l\kappa_{\text{bb},\,l}+\sum_{i}\left(\kappa_{\text{bf},\,i}+\kappa_{\text{ff},\,i}\right).
\end{eqnarray*}
Each set of abundance-specific opacities are saved in a binary file tabulated in wavenumber, density, and temperature. The opacities are interpolated in density and temperature for each individual wavenumber in the radiative transfer calculations using two-dimensional rational splines \citep{Sp:95}.

We were at first kindly provided with the same opacity table for solar metallicities that {\rBl} use (Aringer, priv. comm.). Due to unknown reasons, we were unsuccessful in using these data with our new models. We discuss these extant opacity data and make a simple comparison with our new opacities in Appendix~\ref{app:opaccomp}. 

\begin{table}
\caption{Dust parameters: Forsterite\label{tab:parameters}}
\begin{tabular}{lrlll}\hline\hline\\[-1.5ex]
parameter & value & {\rBl}\tablefootmark{a} & unit & Reference\\[1.0ex]\hline\\[-1.8ex]
$\epsilon_{\text{Fo}}$ & $10^{-15}$ & $10^{-15}$ & & 1\\
$\mathcal{A}_{\text{Fo}}$ & 140.694 & 140 & & 2, 3\\
$\rhoint{\text{Fo}}$ & 3.21 & 3.27 & $\text{g}\,\text{cm}^{-3}$ & 2, 3\\
$\xi_{\text{Fo}}$ & 0.1 & 1.0 & & 4\\
$E_{\text{b,Fo}}$ & 3.5 & -- & eV & 5\\
$\nu_{\text{Fo}}^{\text{Si}}$, $\nu_{\text{Fo}}^{\text{Mg}}$ & 1, 2\tablefootmark{b} & 1, --\\[1.0ex]\hline
\end{tabular}
\tablebib{(1)~{\rBl}; (2)~\citet{Li:95}; (3)~{\rGS}, Table~12.1; (4)~\rGS, Sect.~12.7.1; (5)~\citet{Ba:78}, the `Silicate' entry in Table~4}
\tablefoot{
\tablefootmark{a}{The values we assume {\rBl} use are specified by \citet{HoBlArAh:16}.}
\tablefoottext{b}{For all elements $j$, but Mg and Si, $\nu_{\text{Fo}}^{j}=0$.}
}
\end{table}

\begin{figure}
    \centering
    \includegraphics[]{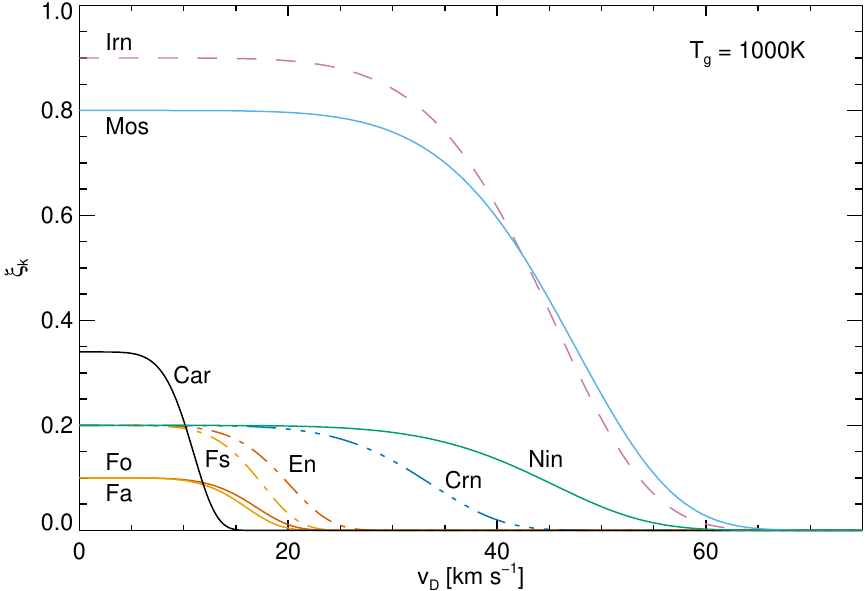}
    \caption{Sticking coefficient $\xi_{k}$ (Eq.~(\ref{eq:xik})) vs the drift velocity $v_{\text{D}}$ for nine different minerals, assuming a gas temperature $\Tg=1000\,\text{K}$.}
    \label{fig:xik}
\end{figure}

\subsubsection{Dust properties}\label{sec:propdust}
We list all the forsterite-specific dust parameters we used in Table~\ref{tab:parameters}. We illustrate how the sticking coefficient $\xi_{k}$ varies with the drift velocity for forsterite and eight other minerals in Fig.~\ref{fig:xik}  (cf. Fig.~1 in {\rSaHoc}); the parameters of the eight additional minerals are taken from the same set of references as for forsterite, noting that the binding energies of several minerals are highly uncertain. The figure shows that the sticking coefficient drops to 0 when $\vD\ga20\,\kms$ for larger mineral monomers such as forsterite (Fo), fayalite (Fa), enstatite (En), and ferrosilite (Fs) and also for carbon (Car; amorphous carbon or graphite). Higher velocities are possible with corundum (Crn; $\vD\approx40\,\kms$) as well as iron (Irn), moissanite (Mos), and niningerite (Nin) where the cutoff drift velocity is about $60\,\kms$. Hence, there is for forsterite no grain growth when the drift velocity is higher than $20\,\kms$.

All the models were calculated using Mie scattering. Optical constants $(n_{\nu},\,k_{\nu})$ are   from \citet{JaDoMu.:03}.\footnote{The optical data can be retrieved from \href{https://www.astro.uni-jena.de/Laboratory/OCDB/amsilicates.html}{https://www.astro.uni-jena.de/Laboratory/OCDB/amsilicates.html}.} 
To achieve results comparable to those of {\rBl}, we used the seed particle abundance $\epsilon_{\text{Fo}}=10^{-15}$ with all our calculations, as well as the same sticking coefficient with our PC models, $\xi^{(\text{Fo})}( \text{PC})=1.0$.\footnote{For the  sticking coefficients we use $\xi^{(\text{Fo})}=1$ with our PC models to allow a direct comparison with previous works regardless of the selected parameters. However, $\xi^{(\text{Fo})}=1$ is not realistic. The assumption was introduced in the design of the model grid project of \citet{MaWaHo:10} to compensate for too low rates of grain growth for many combinations of stellar parameters, which is a problem related to the PC assumption. When drift is added, the $\xi^{(\text{Fo})}=1$ assumption is no longer necessary and a value near unity is more challenging in drift models, which can become unstable; plausibly, grains grow too fast in the inner wind region. The assumption $\xi^{(\text{Fo})}=0.1$ is more realistic and does not seem to result in unstable models. We calculated one drift model (L3.85T24) with both $\xi^{(\text{Fo})}=0.1$ and $\xi^{(\text{Fo})}=1$ to demonstrate the role of the sticking coefficient in a scenario that accounts for drift.}

\begin{table}
\caption{Model parameters. The six columns specify the model set-up name, stellar mass {\Ms}, stellar luminosity {\Ls}, effective temperature {\teff}, pulsation amplitude $\deltauvelp$, and luminosity-dependent pulsation period $P$.}
\label{tab:resmodpar}
\begin{tabular}{lccccr}\hline\hline\\[-1.8ex]
model & \Ms &$\log\left(\Ls\right)$       &{\teff}     & \deltauvelp & \multicolumn{1}{c}{\pP}\\
      &$[\Msun]$ &$[\Lsun]$&$[\mbox{K}]$& $[\kms]$ & \multicolumn{1}{c}{$[\mbox{d}]$}\\[1.0ex]\hline\\[-1.8ex]
L3.85T24 & 1.00 & 3.85 & 2400 & 3.0 & 478\\
L3.85T27 & 1.00 & 3.85 & 2700 & 4.0 & 478\\
L4.30T29 & 1.00 & 4.30 & 2900 & 4.0 & 1060\\
L4.45T29 & 1.50 & 4.45 & 2900 & 4.0 & 1376\\[1.0ex]\hline
\end{tabular}
\end{table}

\subsubsection{Selection of model parameters}\label{sec:selmodel}
Our new model calculations of M-star objects are as demanding as those of C stars in {\rSaMa}. The initial results of our new M-type wind models   indicated drastically lower mass-loss rates in our current forsterite-driven drift models, which is why we found it necessary to select stellar parameter sets similar to non-drift simulations where results are known to give high mass-loss rates. Here we calculated four sets of models to get a first impression of how the formation of stellar winds of M stars work when the dust component is allowed to drift relative to the gas. All model parameters are collected in Table~\ref{tab:resmodpar}.

Our focus was to vary the temperature and luminosity; this also sets the period (see Sect.~\ref{sec:modelpar}). We used a higher value with the piston amplitude as such a value is more likely to result in the formation of a stellar wind using drift. There is little reason to fine-tune this amplitude, given all the  other  still  imprecise factors of these new models. Finally, we used the same value on the stellar mass, $\Ms=1\,\Msun$, with the exception of the model with the highest luminosity where we used $\Ms=1.5\,\Msun$ (see below).

Firstly, we selected the proof-of-concept model set B of {\rHo} ($\log(L/\Lsun)=3.85$, $\teff=2700\,\text{K}$), which she uses to illustrate that dust-driven winds also form in M-type stars when Mie scattering is used in the description of dust extinction instead of the SPL approximation. Secondly, we selected a model set of {\rBl} with a lower effective temperature that is also found to form a higher mass-loss rate ($\log(L/\Lsun)=3.85$, $\teff=2400\,\text{K}$). 

\citet{MaDeKa:23} consider a small sample of AGB stars,  observed with the Atacama Large Millimeter Array (ALMA),  that show spectral energy distributions of amorphous silicates characterised by deep absorption features at 10 and 18 $\mu$m.  Based on their observed properties and evolutionary tracks, we find it valuable to also consider a higher effective temperature using a lower and higher luminosity. Consequently, we added a third set with $\log(L/\Lsun)=4.30$, $\teff=2900\,\text{K}$, and $\Ms=1.0\,\Msun$, as well as a fourth set with $\log(L/\Lsun)=4.45$, $\teff=2900\,\text{K}$, and $\Ms=1.5\,\Msun$. The last set-up corresponds, roughly, to the late-stage evolution beyond the mass-loss peak in the evolutionary track with initial mass $M=6\,\Msun$ in \citet{MaDeKa:23}. The third set-up is a compromise to capture mass-loss characteristics of the $M=4\,\Msun$ and $M=5\,\Msun$ tracks.

\begin{sidewaystable*}
  \caption{Temporally averaged quantities at the outer boundary (see Section~\ref{sec:results}). From the left, the first three columns specify the model name (see Table~\ref{tab:resmodpar}), if PC or drift is used (P/D), and the sticking coefficient ($\xi^{(\text{Fo})}$). The following nine columns  show the temporally averaged values of mass-loss rate {\mmdot}, terminal velocity {\muinf}, terminal drift velocity {\mvdrinf}, modified {\mfcondsim} and true {\mfcondsi} degrees of condensation of silicon, modified {\mfcondmgm} and true {\mfcondmg} degrees of condensation of magnesium, dust-to-gas mass-loss ratio {\mdrhog}, and dust radius {\mradd{Fo}}. A relative fluctuation amplitude {\rfluc} is provided for each property; the subscript $m$ indicates that the shown value was multiplied by a factor of $10^3$. All values are indented to make visible the large variation in values of each property across all models. The final columns show the outflow classification class: periodic ($l\times$p) and quasi-periodic ($l$q); $l$ indicates the (multi-)periodicity of the gas and dust outflow in the unit of the piston period $P$. The rows of drift models are shown in boldface.}
\label{tab:resall}
\begin{tabular}{$l@{\quad}^c@{\:\:}c@{\:\:}^r%
                @{\ }^l@{\:\:}^ll%
                @{\ }^l@{\:\:}^ll%
                @{\ }^l@{\:\:}^ll%
                @{\ }^l@{\:\:}^ll%
                @{\ }^l@{\:\:}^ll%
                @{\ }^l@{\:\:}^ll%
                @{\ }^l@{\:\:}^ll%
                @{\ }^l@{\:\:}^ll%
                @{\ }^l@{\:\:}^l^r}\hline\hline\\[-1.8ex]
   \multicolumn{1}{l}{model} & P/D & $\xi^{(\text{Fo})}$ &&
        \multicolumn{2}{c}{$10^8$\,\mmdot} &&
        \multicolumn{2}{c}{\muinf}  &&
        \multicolumn{2}{c}{\mvdrinf}  &&
        \multicolumn{2}{c}{\mfcondsim} &&
        \multicolumn{2}{c}{\mfcondsi} &&
        \multicolumn{2}{c}{\mfcondmgm} &&
        \multicolumn{2}{c}{\mfcondmg} &&
        \multicolumn{2}{c}{$10^4$\mdrhogf} &&
        \multicolumn{2}{c}{$10^2$\mradd{Fo}} & cl.\\
& &&& \multicolumn{2}{c}{$[\mdotu]$} &&
        \multicolumn{2}{c}{$[\kms]$}    &&
        \multicolumn{2}{c}{$[\kms]$}    &&
        \multicolumn{2}{c}{}      &&
        \multicolumn{2}{c}{}      &&
        \multicolumn{2}{c}{}      &&
        \multicolumn{2}{c}{}      &&
        \multicolumn{2}{c}{} &&
        \multicolumn{2}{c}{$[\mu\text{m}]$}      \\
& &&&   &\multicolumn{1}{c}{\rfluc} &&
        &\multicolumn{1}{c}{\rfluc} &&
        &\multicolumn{1}{c}{\rfluc} &&
        &\multicolumn{1}{c}{\rfluc} &&
        &\multicolumn{1}{c}{\rfluc} &&
        &\multicolumn{1}{c}{\rfluc} &&
        &\multicolumn{1}{c}{\rfluc} &&
        &\multicolumn{1}{c}{\rfluc} &&
        &\multicolumn{1}{c}{\rfluc}\\[1.0ex]\hline\\[-1.0ex]
%                      PD   u_inf                              Mdot                    f_cond                    r_d                                               dmd                                  vdr              comment
L3.85T24 & P & 1.0
&& 218 & \textit{$\phantom{0}$24}
&& 11.1 & \textit{$\phantom{0}$0.17}
&& &
&& $\phantom{0}$0.184 & \textit{40}$_{m}$ && $\phantom{0}$0.184 & \textit{40}$_{m}$
&& 0.344 & \textit{74}$_{m}$ && $\phantom{0}$0.344 & \textit{74}$_{m}$
&& $\phantom{00}$6.62 & \textit{$\phantom{00}$1.4}
&& 48.2 & \textit{$\phantom{0}$3.6} & 3q\\[0.15ex]
         & P & 0.1
&& $\phantom{0}$32.8 & \textit{$\phantom{00}$0.38}
&& $\phantom{0}$4.89 & \textit{42}$_{m}$
&& &
&& $\phantom{0}$0.127 & $\phantom{0}$\textit{3.0}$_{m}$
&& $\phantom{0}$0.127 & $\phantom{0}$\textit{3.0}$_{m}$
&& 0.238 & $\phantom{0}$\textit{5.7}$_{m}$ 
&& $\phantom{0}$0.238 & $\phantom{0}$\textit{5.7}$_{m}$
&& $\phantom{00}$4.58 & \textit{$\phantom{00}$0.11}
&& 42.8 & \textit{$\phantom{0}$0.34} & 1q\\[0.15ex]
\:{\rBl} & P & 1.0
&& 248 &
&& 10.0 &
&& &
&& $\phantom{0}$0.18 &
&& $\phantom{0}$0.18 &
&& &
&& &
&& $\phantom{00}$6.49 &
&& 48 & &--\\[0.15ex]
\rowstyle{\bfseries} & D & \textbf{1.0}
&& $\phantom{00}$1.6 & \textit{$\phantom{00}$0.16}
&& 13.3 & \textit{$\phantom{0}$0.16}
&& 297 & \textit{19}
&& $\phantom{0}$0.327 & $\phantom{0}$\textit{0.12}
&& 14.0$_{m}$ & $\phantom{0}$\textit{4.8}$_{m}$
&& 0.611 & $\phantom{0}$\textit{0.22}
&& 26.1$_{m}$ & $\phantom{0}$\textit{8.9}$_{m}$
&& 278 & \textit{110}
&& 57.9 & \textit{$\phantom{0}$6.3}  & 1p\\[0.15ex]
\rowstyle{\bfseries} & D & \textbf{0.1}
&& $\phantom{00}$3.8 & \textit{$\phantom{00}$0.38}
&& 13.1 & \textit{$\phantom{0}$0.15}
&& 194 & \textit{15}
&& $\phantom{0}$0.294 & $\phantom{0}$\textit{0.18}
&& 18.2$_{m}$ & $\phantom{0}$\textit{9.6}$_{m}$
&& 0.549 & $\phantom{0}$\textit{0.33}
&& 34.0$_{m}$ & \textit{18}$_{m}$
&& 173 & \textit{117}
&& 55.0 & \textit{$\phantom{0}$9.2}  & q\\[0.15ex]
L3.85T27 & P & 1.0
&& $\phantom{0}$19.4 & \textit{$\phantom{00}$2.7}
&& $\phantom{0}$4.64 & \textit{80}$_{m}$
&& &
&& $\phantom{0}$0.125 & \textit{11}$_{m}$
&& $\phantom{0}$0.125 & \textit{11}$_{m}$
&& 0.233 & \textit{21}$_{m}$
&& $\phantom{0}$0.233 & \textit{21}$_{m}$
&& $\phantom{00}$4.48 & \textit{$\phantom{00}$0.40}
&& 42.5 & \textit{$\phantom{0}$1.2} & 1q\\[0.15ex]
\:\rHo-B & P & 1.0
&& $\phantom{0}$80 &
&& $\phantom{0}$7 &
&& &
&& $\phantom{0}$0.15 &
&& $\phantom{0}$0.15 &
&& &
&& &
&& &
&& 45 & &--\\[0.15ex]
\:{\rBl} & P & 1.0
&& 362 &
&& $\phantom{0}$9.2 &
&& &
&& $\phantom{0}$0.15 &
&& $\phantom{0}$0.15 &
&& &
&& &
&& $\phantom{00}$5.20 &
&& 45 & &--\\[0.15ex]
\rowstyle{\bfseries} & D & \textbf{0.1}
&& $\phantom{00}$1.21 & \textit{$\phantom{00}$0.21}
&& 11.9 & \textit{$\phantom{0}$0.54}
&& 313 & \textit{34}
&& $\phantom{0}$0.290 & $\phantom{0}$\textit{0.17}
&& 10.5$_{m}$ & $\phantom{0}$\textit{5.9}$_{m}$
&& 0.542 & $\phantom{0}$\textit{0.33}
&& 19.5$_{m}$ & \textit{10.9}$_{m}$
&& 296     & \textit{190}
&& 54.3 & \textit{11} & q\\[0.15ex]
L4.30T29 & P & 1.0
&& 663 & \textit{660}
&& $\phantom{0}$7.70 & \textit{$\phantom{0}1.3$}
&& &
&& 88.7$_{m}$ & \textit{46}$_{m}$
&& 88.7$_{m}$ & \textit{46}$_{m}$
&& 0.165 & \textit{86}$_{m}$
&& $\phantom{0}$0.165 & \textit{86}$_{m}$
&& $\phantom{00}$3.18 & \textit{$\phantom{00}$1.7}
&& 35.8 & \textit{$\phantom{0}$9.5} & q\\[0.15ex]
\rowstyle{\bfseries} & D & \textbf{0.1}
&& 631 & \textit{180}
&& $\phantom{0}$7.83 & \textit{$\phantom{0}$0.49}
&& $\phantom{0}$13.0 & \textit{$\phantom{0}$4.7}
&& 77.0$_{m}$ & \textit{85}$_{m}$
&& 25.4$_{m}$ & \textit{24}$_{m}$
&& 0.144 & \textit{$\phantom{0}$0.16}
&& 47.5$_{m}$ & \textit{45}$_{m}$
&& $\phantom{00}$8.76 & \textit{$\phantom{0}$11}
&& 32.6 & \textit{11} & q\\[0.15ex]
L4.45T29 & P & 1.0
&& 278 & \textit{$\phantom{0}$31}
&& $\phantom{0}$5.17 & \textit{$\phantom{0}$0.12}
&& &
&& 56.7$_{m}$ & \textit{$\phantom{0}$5.3}$_{m}$
&& 56.7$_{m}$ & \textit{$\phantom{0}$5.3}$_{m}$
&& 0.106 & \textit{$\phantom{0}$9.9}$_{m}$
&& $\phantom{0}$0.106 & \textit{$\phantom{0}$9.9}$_{m}$
&& $\phantom{00}$2.04 & \textit{$\phantom{00}$0.19}
&& 32.7 & \textit{$\phantom{0}$1.0} & i\\[0.15ex]
\rowstyle{\bfseries} & D & \textbf{0.1}
&& $\phantom{0}$35.5 & \textit{$\phantom{00}$2.6}
&& $\phantom{0}$9.87 & \textit{92}$_{m}$
&& $\phantom{0}$86.7 & \textit{21}
&& 89.3$_{m}$ & $\phantom{0}$\textit{54}$_{m}$
&& $\phantom{0}$8.66$_{m}$ & $\phantom{0}$\textit{3.1}$_{m}$
&& 0.167 & $\phantom{0}$\textit{0.10}
&& 16.2$_{m}$ & \textit{$\phantom{0}$5.9}$_{m}$
&& $\phantom{0}$35.1 & \textit{$\phantom{0}30$}
&& 36.8 & \textit{$\phantom{0}$6.5} & 1p\\[1.0ex]\hline\\[-1.0ex]
\end{tabular}
\end{sidewaystable*}

\subsection{Results}\label{sec:results}
As in {\rSaMa}, we characterise wind models with a set of properties that are temporally averaged at the outer boundary. The mass-loss rate {\mmdot} and the terminal velocity {\muinf} characterise the gas. The degree of condensation of silicon {\mfcondsi} and magnesium {\mfcondmg}, the dust-to-gas mass-loss ratio {\mmdmdot}, the mean grain radius {\mdrad}, and the terminal drift velocity {\mvDinf} characterise the dust.

The drift-velocity dependent true degree of condensation of tracer element $j$ accounting for all minerals $k$ is calculated as follows (cf. Eq.~(12.81) in {\rGS}\footnote{We ignore the seed particle radius $a_{0,j}$ used by {\rGS};  the contribution of seed particles to the degree of condensation is minuscule already at small particle radii $a_{j}$, which are barely larger than $a_{0,j}$.} and Eq.~(33) in {\rSaMa}):
\begin{eqnarray}
f_{\text{cond}}^{j}=\sum_k\frac{4\pi a_{j}^3}{3}\frac{1}{V_{1,j}}\frac{1}{\FDk}\frac{\epsilon_{k}\nu_{k}^{j}}{\varepsilon_{j}}=\frac{1}{\varepsilon_{j}}\sum_k\frac{N_{k}\epsilon_{k}\nu_{k}^{j}}{\FDk}.
\end{eqnarray}
Here $\FDk=1+\left|v_{\text{D},k}\right|/\left(\left|u\right|+\varepsilon_u\right)$ is the drift factor,\footnote{This form of the drift factor allows calculation of the degree of condensation also when the sign of the gas and dust velocity components differ, which could be the case in the radial region where the stellar wind first forms in time-dependent simulations.} where we use $\varepsilon_u=0.1\,\kms$. The variable $\varepsilon_{j}$ is the abundance of element $j$. In PC models, the modified degree of condensation is (cf. Eq.~(12.85) in {\rGS})
\begin{eqnarray}
\widetilde{f}_{\text{cond}}^{\,\,j}=\frac{1}{\varepsilon_{j}}\sum_kN_{k}\epsilon_{k}\nu_{k}^{j}.
\end{eqnarray}
Whilst the true degree of condensation corresponds to the modified degree of condensation diluted by the relative velocity of dust to gas, we also present modified degrees of condensation for drift models as these values illustrate the more efficient dust formation process in drift models. Moreover, the dust-to-mass mass-loss ratio is
\begin{eqnarray}
\frac{\dmdot}{\mdot}=\sum_k\frac{\rhodk}{\rhog}\frac{\vinfk}{\uinf}=\sum_k\ddgk\FDki,
\end{eqnarray}
where $\ddgk=\rhodk/\rhog$.

We calculated a relative fluctuation amplitude $\hat{r}$ for each property $\mathcal{Q}$ as $\hat{r}=\sigma_{\text{s}}/\mathcal{Q}$, where $\sigma_{\text{s}}$ is the (sample) standard deviation of the property $\mathcal{Q}$ in the time interval that is used to measure the same property. We show results of our model calculations in Table~\ref{tab:resall}.

\section{Discussion}\label{sec:discussion}
We analyse the new results of our PC models in Sect.~\ref{sec:discussPC} and then compare PC models with our new drift models in Sect.~\ref{sec:discussdrift}.

\subsection{Comparing results of the PC (non-drift) models}\label{sec:discussPC}

\subsubsection{Lower-temperature model L3.85T24}
Our PC models using a unity sticking coefficient ($\xi^{\text{(Fo)}}=1.0$) reveal a more massive wind than when $\xi^{\text{(Fo)}}=0.1$. The mass-loss rate and expansion velocity are 5.6 and 1.3 times higher, respectively. A lower ratio of 1.4 is seen in the degree of condensation and the dust-to-gas density ratio. The average grain radius is 13\,\% higher. Fluctuation amplitudes are 3--12 times higher in all values, except for the mass-loss rate, where the amplitude is 62 times higher. Clearly, a unity sticking coefficient results in a more variable wind where all values except the grain radius are higher, but the increase is with the exception of the mass-loss rate of 13--130\,\% and not a factor of 10.

In comparison to {\rBl}, and assuming a unity sticking coefficient, our values on the degrees of condensation and dust-to-gas density ratio are 2.0--2.2\,\% higher, the grain radius 0.42\,\% higher, and the mass-loss rate 12\,\% lower. These values compare well.

\subsubsection{Proof-of-concept model L3.85T27}
\citet{Ho:08}
presents model L3.85T27 (`model $B$') as a proof of the concept that scattering on larger dust particles in place of absorption allows formation of massive stellar winds in M-type stars. {\rBl} calculate a model with the same stellar parameters, also with 100 grid points, but use a somewhat longer pulsation period ($P=478\,$d, instead of $P=390\,$d) and also favour a smaller pulsation amplitude ($\deltauvelp=2\,\kms$). We used the same stellar parameters as {\rBl}, where $\deltauvelp=4\,\kms$. We also set the sticking coefficient to unity, but we used a higher spatial resolution achieved with $N_{\text{D}}=840$ grid points. We show the resulting values of both {\rHo} and {\rBl} in Table~\ref{tab:resall} for easy reference.

Our new values on the silicon degree of condensation and the mean grain radius agree well with these two studies; our values are 83\,\% and 94\,\% of their values, respectively. Similarly, our value on the dust-to-gas ratio is 86\,\% of the value of {\rBl}. Our values for the mass-loss rate and expansion velocity  are 24\,\% and 66\,\% (5.4\,\% and 50\,\%) of the values of {\rHo} ({\rBl}). Our mass-loss rate is closer to the value of {\rHo}, whilst the difference is much larger in comparison to {\rBl}. Notably, the agreement with {\rBl} is not as close as in model L3.85T24. In any case, the agreement depends on where exactly the inner boundary is located as that determines the amount of mass in the model domain. As neither {\rBl} nor other authors specify the location of the inner boundary or the mass in the model domain, there is some arbitrariness to the agreement of results; this also applies to model L3.85T24 above.

\subsubsection{Two higher-temperature models: L4.30T29 and L4.45T29}
Model L4.30T29 shows the highest mass-loss rate of all four PC models, with associated large variabilities. The expansion velocity remains low. The degree of condensation is about half the value of models L3.85T24 and L3.85T27; this is also reflected in the dust-to-gas mass-loss ratio. The mean grain size is somewhat lower than in the first two models.

Model L4.45T29 is modelled using both a higher mass and a higher luminosity than L4.30T29. The results show a mass-loss rate that is  58\,\% lower. The remaining properties of the more luminous model are 33--36\,\% lower. The exception is the average grain size, which is 8.7\,\% lower. Compared to the same model, relative fluctuation amplitudes of the mass-loss rate are 4.6\,\% and of the remaining properties 9.2--12\,\%. It is clear that the higher luminosity does not compensate for the higher gravitational pull owing to the higher stellar mass. It is noteworthy that both luminous model set-ups result in high mass-loss rates, whilst forming smaller amounts of dust than in models L3.85T24 and L3.85T27.

\subsection{Comparing results of the drift models}\label{sec:discussdrift}

\subsubsection{L3.85T24: Effects of the sticking coefficient}
The drift model shows a different result in comparison to the PC model L3.85T24. The values of the two drift models are similar when comparing properties that do not depend on the dust velocity (they differ by 5.3--15\,\%), with the exception of the mass-loss rate where the value of the unity sticking coefficient model is 42\,\% lower than in the other model. Due to the 53\,\% higher drift velocity of the unity sticking coefficient model, the dust-to-gas mass-loss rate ratio differs by 61\,\%, and the true degrees of condensation by $-23$\,\%. Thus, the model with the lower sticking coefficient results in a much higher mass loss. A plausible explanation to this is that less efficiently formed dust grains can flow to regions where they more effectively contribute to wind formation before there is enough dust to accelerate both the dust and gas outwards. Notably, the very high drift velocity of 194\,{\kms} increases by 53\% to 297\,{\kms} in the unity model. The true degrees of condensation are a factor of  15--22 lower than the modified degrees of condensation owing to the high drift velocities. The fluctuation amplitudes are, finally, all very similar (they differ by $-58$--27\,\%).

We compare the drift model using $\xi^{\text{(Fo)}}=0.1$ with our PC model, where $\xi^{\text{(Fo)}}=1.0$. The mass-loss rate of the drift model is 1.7\,\%, the dust-to-gas density ratio 2600\,\%, and the true degrees of condensation 9.9\,\% of the corresponding values of the PC model. The differences are much smaller in the expansion velocity (18\,\% higher) and the grain radius (14\,\% higher). The same comparison with the  {\rBl} model gives a mass-loss rate that is 1.5\,\%, a dust-to-gas density ratio that is 2700\,\%, and a degree of condensation of silicon 10\,\% of their values. Similarly, the differences are smaller in the expansion velocity (30\,\% higher) and the grain radius (15\,\% higher). The huge increase in the dust-to-gas density ratio of the drift model must be put in context of the drastically lower mass-loss rate; dust forms more efficiently, whilst the wind formation efficiency is lower. Nevertheless, the differences in comparison to the PC models of about a factor of 1/50 in the mass-loss rate and a factor of 26 in the dust-to-gas density ratio are large and must not be ignored.

\begin{figure*}
\centering
\includegraphics[width=\textwidth]{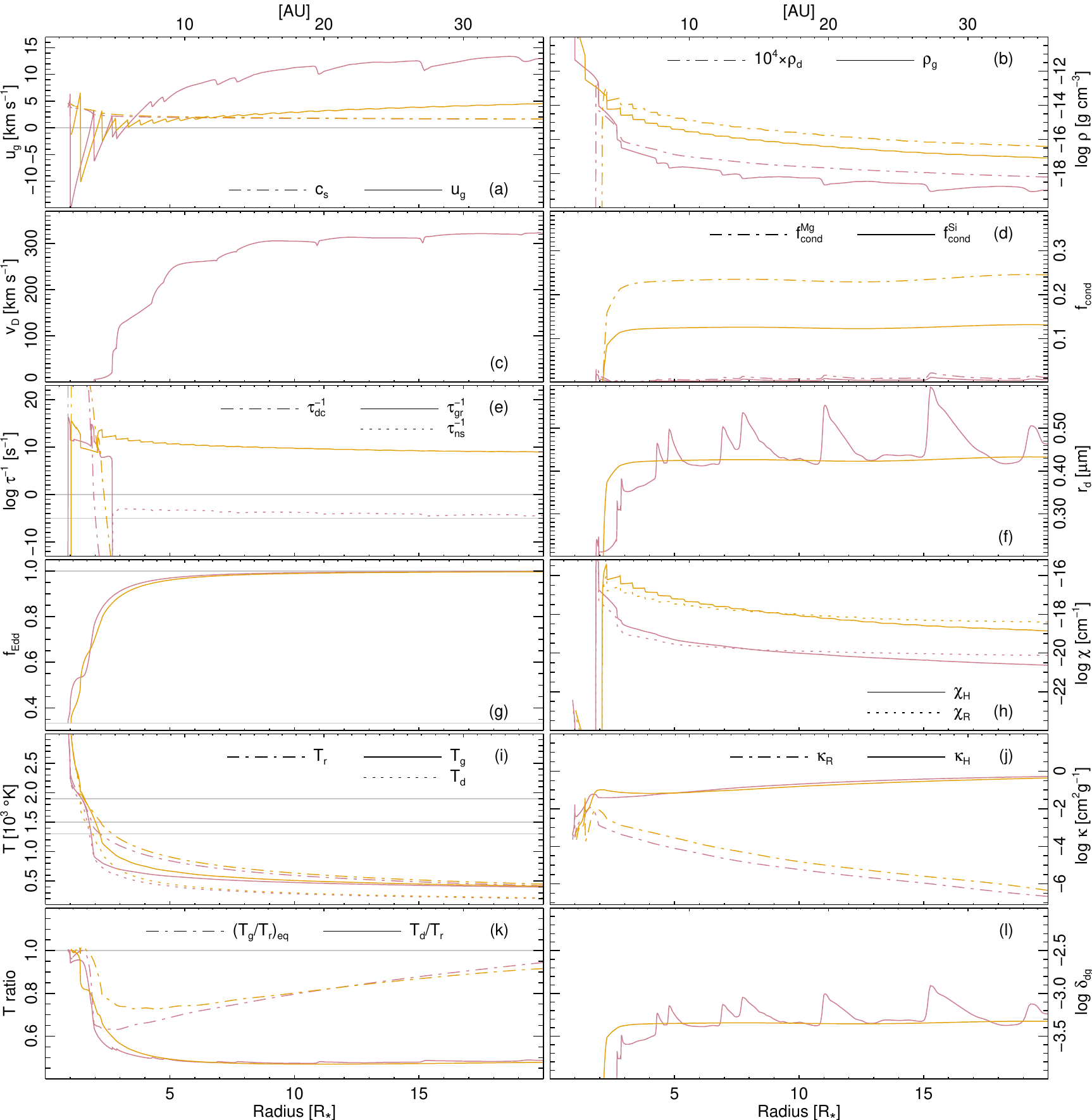}
\caption{Radial structure of  snapshot of set-up L3.85T27 for the full modelled region. Shown are the drift model, where $\xi^{(\text{Fo})}=0.1$ (purple), and  the PC model, where $\xi^{(\text{Fo})}=1.0$ (orange). From the top left, the 12 panels show: (a) gas velocity $u_{\text{g}}$, sound speed $\cs$; (b) gas density $\rhog$, dust density $10^4\times\rhod$ (log); (c) drift velocity $\vD$; (d) true degree of condensation of silicon $\fconde{Si}$ and magnesium $\fconde{Mg}$; (e) net growth rate {\tauGi}, net decay rate {\taudc}, and  non-thermal sputtering {\tauns} (log) times $n_{\text{d,Fo}}\text{d}V$ (where d$V$ is the cell volume); (f) average grain radius {\radd}; (g) Eddington factor {\fedd}; (h) extinction coefficient {\chiH}, Rosseland mean extinction coefficient {\chiR}; (i) gas temperature {\Tg}, radiative temperature {\Tr}, and dust temperature {\Td}; (j) gas opacity {\kappaH} and Rosseland mean opacity {\kappaR} (log); (k) temperature ratios ${\Td}/{\Tr}$ and $(\Tg/\Tr)_{\text{eq}}$; and (l) dust-to-gas density ratio {\ddg}. All properties are drawn vs the stellar radius $R_{*}$ (lower axis) and astronomical units (AU; upper axis). The grey horizontal lines are guides.}\label{fig:poc_radial}
\end{figure*}

\subsubsection{Proof-of-concept drift model L3.85T27}
The values of the drift model again differ from our PC model. The mass-loss rate is here 94\,\% lower, the dust-to-gas density ratio 65 times higher, and the degrees of condensation 92\,\% lower. Furthermore, the expansion velocity is 160\,\% higher, and the mean grain radius 28\,\% higher. Temporal fluctuations are higher in the dusty properties, but the fluctuations are generally low. Both the drift model and the PC model show quasi periodic variations. The mass-loss rate is 0.33\,\% of the value of {\rBl}, the dust-to-gas density ratio 56 times higher, and the degree of condensation of silicon is 7\,\% of the value of {\rBl}. The expansion velocity is 29\,\% higher and the grain radius 21\,\% higher.

We plot our PC and drift models versus the radius in Fig.~\ref{fig:poc_radial}. The inefficient mass loss of the drift model is seen in the gas density, which  is a hundred times lower than in the PC model (Fig.~\ref{fig:poc_radial}b). Despite the lower mass-loss rate, the expansion velocity of the drift model is more than twice as high than in the PC model (see Fig.~\ref{fig:poc_radial}a).  The gas opacity {\kappaH} increases somewhat towards higher radii (Fig.~\ref{fig:poc_radial}j), whilst the Rosseland mean opacity {\kappaR} decreases to be a millionth or less of {\kappaH} at higher radii; notably, the same ratio is closer to a factor of ten in the carbon-rich model shown in Fig.~8j in {\rSaMa}. The Eddington factor of the two models (Fig.~\ref{fig:poc_radial}g) indicates that there is no need to solve the equation of radiative transfer where $R\ga10\,\Rs$ as the factor is very close to unity.

The drift velocity shown in Fig.~\ref{fig:poc_radial}c attains high values already near the star. The same high values cause an abrupt cutoff in the dust formation rates at radii $r\ga3\,\Rs$ (Fig.~\ref{fig:poc_radial}e). The same panel illustrates that, despite the high drift velocity, non-thermal sputtering ($\tau_{\text{sp,n}}^{-1}$) is negligible at all radii owing to the low gas density. The true degree of condensation (Fig.~\ref{fig:poc_radial}d) and extinction coefficient (Fig.~\ref{fig:poc_radial}h) are significantly lower in the drift model.

The dust-to-radiative temperature ratio in Fig.~\ref{fig:poc_radial}k shows a value that is the inverse of the same ratio in a stellar wind of a carbon star (cf. Fig.~8k in {\rSaMa}). The forsterite dust temperature is always lower than the radiative temperature; it is about half as high in the outer regions. In comparison, the amorphous carbon dust temperature is always higher than the radiative temperature; and it is up to about 50\,\% higher in the outer regions. No other indicator illustrates the difference as clearly between scattering and absorption dominated dust extinction. 

\subsubsection{Two higher-temperature models: L4.30T29 and L4.45T29}
The drift and PC models L4.30T29 show results that are very similar. This similarity is owing to the drift velocity, $\vD=13\,\kms$, which is drastically lower than in the other three drift models of this study. Comparing the other values, the mass-loss rate is 4.8\,\% lower, the expansion velocity 1.7\,\% higher, the modified degrees of condensation 13\,\% lower, the true degrees of condensation 71\,\% lower, the dust-to-gas mass-loss ratio 180\,\% higher, and the average grain size 8.9\,\% lower. The relative fluctuation amplitude of the dust-to-gas mass-loss ratio and average grain size are 550\,\% and 16\,\% higher, respectively. The same values of the modified degrees of condensation are 85--86\,\% higher. The fluctuation amplitudes of all remaining properties are 48--73\,\% lower.

This model set-up shows a similarly strong stellar wind for an M-type star as the models of the C-type star IRC\,10$^\circ$216 discussed in \citet{AnLoMe.:22}, which is modelled using an even less massive star. In addition, of the models discussed here, this is the only set-up where a   smaller amount of forsterite is able to drive a more massive wind.

Model L4.45T29 shows a drift velocity that is 5.7 times higher compared to model L4.30T29, and the mass-loss rate of the former model is 5.6\,\% of the latter. Comparing the values of the drift model with those of the PC model, the mass-loss rate is 13\,\%, the expansion velocity 91\,\% higher, the modified degrees of condensation 57\,\% higher, the true degrees of condensation 58\,\% lower, the dust-to-gas mass-loss ratio 16 times higher, and the average grain size 13\,\% higher. The relative fluctuation amplitude of the dust-to-gas mass-loss ratio and average grain size are 157 and 5.5 times higher, respectively. The same values of the modified degrees of condensation are nine times higher. The fluctuation amplitude of the mass-loss rate is 8.4\,\% of the PC model value;   the fluctuation amplitude of the expansion velocity and the true degrees of condensation are 23\,\% and 40--42\,\% lower.

As in the PC model, the higher stellar mass of this set-up cancels the effect of the higher luminosity. In agreement with model L4.30T29 and despite a smaller amount of dust, this model also forms a more massive wind than do models L3.85T24 and L3.85T27.

\subsubsection{Implications of allowing gas-to-dust drift}
For all four of the  model set-ups presented, the measured expansion velocity of the drift model is higher than that of the corresponding PC model. The mass-loss rate of the drift model is, at the same time, lower to drastically lower, which could be taken as a sign of stellar wind formation in an O-rich chemistry of an M-type star. In the models of C-type stars in {\rSaMa}, the same result holds for only 6 out of 22 model set-ups. For example, the gas density of model set-ups L3.85T24 and L3.85T27 is lower by about a factor of 100 in the drift model compared to the PC model (see Fig.~\ref{fig:poc_radial}b for the latter set-up). In these two cases of extremely low drift model gas densities and mass-loss rates, it is clear that the expansion velocity of the drift models can be higher based on the argument that it is easier to accelerate the gas (and thereby the wind) when the gas density is lower. However, when mass-loss rates of drift and PC models using the same model set-up are more similar, there is no unambiguous evidence for this being a rule that applies to all winds of M-type stars. For example, the differences in expansion velocities and mass-loss rates is smaller with model set-up L4.30T29.

The differences between results of the four drift models and the corresponding PC models presented here are with one exception enormous. Our values on the mass-loss rates of the PC models are 56, 15, and 6.8 times higher than the corresponding drift model. The exception is model L4.30T29 where the difference is only 10\,\%. These values grow to 64 and 300 times when we instead compare them with the corresponding values of {\rBl} for the two less luminous model set-ups. The mass-loss rate of the proof-of-concept model of {\rHo} is 65 times higher than in our drift model. In addition to different model parameters and modelling approach, the differences are owing to the scattering-dominated dust extinction of forsterite, which results in extraordinarily high drift velocities. In the exception model, this does not appear to be the case. Here, with a high luminosity and a lower stellar mass, it is only necessary to form small amounts of dust to drive a massive wind. High drift velocities imply that any features in the diluted dust component in observations will be minuscule. More reliable observations of mass loss based on radio observations of CO, which make fewer assumptions on the dust component, indicate that mass-loss rates can  be high even when the luminosity is less than $10^4\,\Lsun$ \citep[][]{GoOlKe:03};  evidently something is needed to also achieve these high mass-loss rates  in simulations, for example an unidentified component that is not yet included in our models.

The question is what happens when additional dust species are added to the simulations. Not all species are as transparent as forsterite. It seems valuable to study changes in the radiation field and wind driving mechanism where  enstatite is also added, and species that include iron, such as fayalite and ferrosilite, and thus olivine and pyroxene as well as corundum and pure iron dust. \citet{BlHo:12} and  {\rGS} (see Chapter 12) argue that these minerals form much farther out than iron-free minerals, but it is still unknown what the result will be in a multi-fluid model where temperature gradients of drift models may be much steeper in the inner wind-forming region (see Fig.~\ref{fig:poc_radial}i). Drift must not be ignored in stellar winds that are driven by minerals such as forsterite where extinction is dominated by scattering.

\section{Conclusions}\label{sec:conclusions}
We have extended our frequency-dependent dust-driven high-spatial-resolution wind model code {\teh} from {\rSaMa} with descriptions for mineral formation in oxygen-rich chemistry, as laid out by {\rGS}. We have also calculated new opacity tables that are based on bound-bound cross-sections of 30 atoms and molecules of the {\exomol} project and free-free and bound-free opacities of the \textsc{jekyll} code. With our improved model code and opacity data, we can choose molecular compositions and wavelengths freely and model stellar winds of both C-type and M-type stars that form various types of minerals. To our understanding, {\teh} is the physically and numerically most detailed dynamic stellar wind code there is, and it is the only one that can accurately calculate effects of gas-to-dust drift in either type of star.

We have calculated models to explore the effects of drift in M-type winds that are driven by forsterite particles. Extant studies favour this species as a wind driver. We selected model parameter set-ups that are expected to show, and have shown, high mass-loss rates. Our new PC models show a good comparison with extant results (of {\rHo} and {\rBl}); we cannot be more specific as details of those extant simulations are unavailable in the literature.

The differences are much larger when we compare the results of PC models with drift models. Whilst changes in expansion velocities and grain sizes are modest, this is not so for the degree of condensation, dust-to-gas density ratio, and mass-loss rates. The drift velocity is, with one exception, 87--310\,\kms\ in the presented models; these high values result in very low degrees of condensation. One luminous model shows a lower drift velocity of 13\,\kms, which is still about twice as high as the expansion velocity of the same model. The mass-loss rate is 1.7--13\,\% of the PC model value. In the one exception model, the same value is 95\,\%. One of the models showing larger differences (L=7080\Lsun, T=2700K) is important as {\rHo} use a model with the same parameters to prove the concept of stellar wind formation in M-type stars. Drift is more important in M-type stars than in C-type stars;  the biggest difference is that momentum is transferred from the radiation field to the dust grains through scattering on transparent grains instead of through absorption in opaque grains.

More studies are needed that explore the use of simultaneous formation of additional dust species to explain how observed high mass-loss rates form, which  cannot be done without drift;  the resulting simulations are a multi-fluid problem. This article is a proof of concept of the influential effects of drift. Our analysis and results show that effects of drift on stellar wind of M-type AGB stars are strong and that they cannot be correctly determined when drift is ignored.

\begin{acknowledgements}
K.L.C. acknowledges funding from STFC, under project number  ST/V000861/1. Most of the computations were enabled by resources provided by the Swedish National Infrastructure for Computing (SNIC), partially funded by the Swedish Research Council through grant agreement no. 2018-05973.
We thank B. Aringer (Vienna and Padova) for kindly providing us with an opacity table that we could use to develop and test our new model.
\end{acknowledgements}

\bibliographystyle{aa} % style aa.bst
\bibliography{AGB_Refs} % your references Yourfile.bib

\begin{appendix}

\section{Calculation of $\kappa_{\text{bf}}$ and $\kappa_{\text{ff}}$}\label{app:jekyll}
Here we present a brief description of the methods and data used in the determination of the bound-free and free-free opacities, which are calculated using the \textsc{jekyll} code \citep{ErFrJe.:18,ErFr:22}. As in the calculations of our bound-bound opacities, we first determined the cross-sections of bound-free level populations and then calculated opacities based on the physical conditions.

The photo-ionisation cross-sections for ground states are calculated using the analytic fits of \citet{VeYa:95} and \citet{VeFe:96}, and for the lowest excited states of \ion{He}{i}, \ion{C}{i}, \ion{O}{i}, \ion{Mg}{i}, \ion{Mg}{ii}, \ion{Si}{i}, \ion{S}{i}, and \ion{Ca}{ii}, using data from \textsc{TOPbase} of the Opacity Project.\footnote{\href{https://cds.unistra.fr/topbase/topbase.html}{https://cdsweb.unistra.fr/topbase/topbase.html}} For all other excited states, the  photo-ionisation cross-sections are calculated using the hydrogenic approximation by \citet{RyLi:79}. Bound-free opacities are then calculated based on LTE populations of excited and ionised states for given values of density, temperature, and composition. The atomic data (excitation and ionisation energies and statistical weights) were obtained from the Atomic Spectra Database of  NIST\footnote{\href{https://www.nist.gov/pml/atomic-spectra-database}{https://www.nist.gov/pml/atomic-spectra-database}} and the online tables by R. Kurucz.\footnote{\href{https://lweb.cfa.harvard.edu/amp/ampdata/kurucz23/sekur.html}{https://lweb.cfa.harvard.edu/amp/ampdata/kurucz23/sekur.html}}

Free-free opacities $\kappa_{\text{ff},i}$ were calculated for each ion $i$ separately using an expression that depends on the same electron and ion densities  used to calculate bound-free opacities \citep[e.g. Eq.~(5.149) in][]{HuMi:15}.

\section{Role of   gas opacity data}\label{app:opaccomp}
\begin{figure}
\centering
\includegraphics{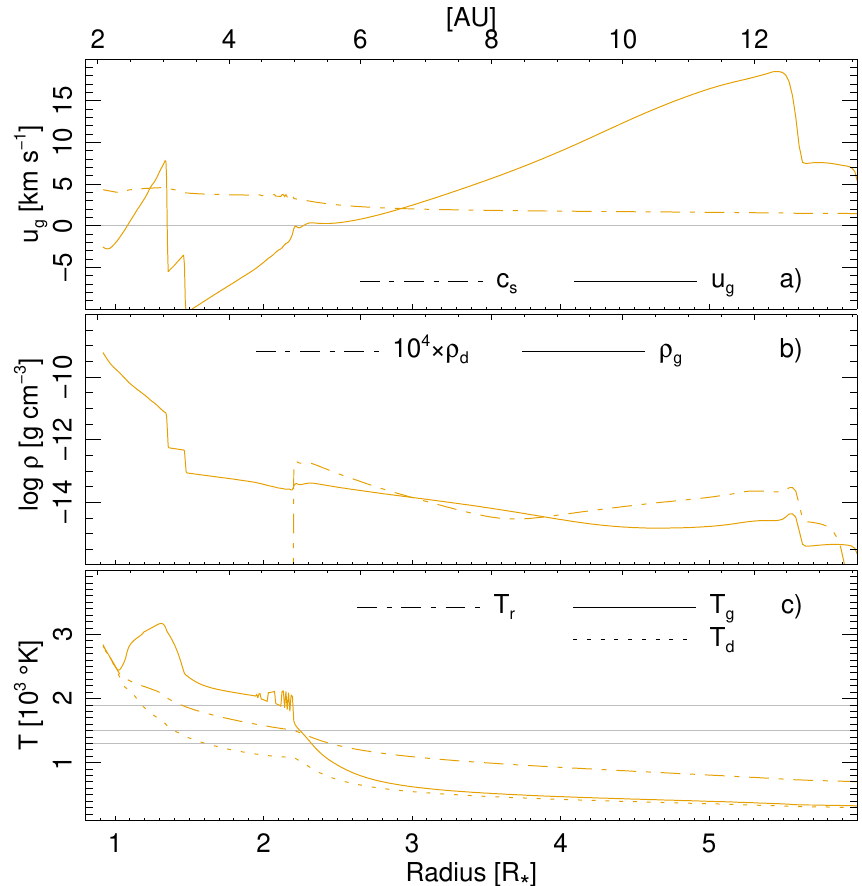}
\caption{Radial structure of  snapshot of set-up L3.85T24 for the inner modelled region. A PC model using the gas opacities of Aringer is shown (in  orange). The three panels show: (a) gas velocity $u_{\text{g}}$, sound speed $\cs$; (b) gas density $\rhog$, dust density $10^4\times\rhod$ (log); and (c) gas temperature {\Tg}, radiative temperature {\Tr}, and dust temperature {\Td}. All properties are drawn vs the stellar radius $R_{*}$ (lower axis) and astronomical units (AU; upper axis). The grey horizontal lines are guides.}\label{fig:optrouble}
\end{figure}
When developing our new oxygen-rich chemistry models, we calculated test wind models using the same gas opacity table that {\rBl} use \citep[the data are described by][]{BAr:00,ArGiNo.:09}. The first test models indicated a problem when starting the stellar wind; some kind of noise appears in the gas temperature of all the models when dust is present and where $\Tg\simeq2000\,$K. We show an example of this noise in Fig.~\ref{fig:optrouble}, which presents a snapshot of a stellar wind model where calculations have just begun. The noise occurs in the temperature in the interval $1.9\la r\la2.2\,\Rs$. The same noise is spatially unresolved in models using $N_{\text{d}}=100$ grid points (not shown), and here the problem is much smaller. The problem becomes prohibitive in drift models, which are more sensitive to variations of this kind. The origin of the noise is unknown, but no terms in the radiation hydrodynamic equations appear to be responsible considering the noise always appears at the same gas temperature. Therefore, we hypothesise  that the noise originates in the gas opacity data.

\begin{figure*}
    \centering
    \includegraphics{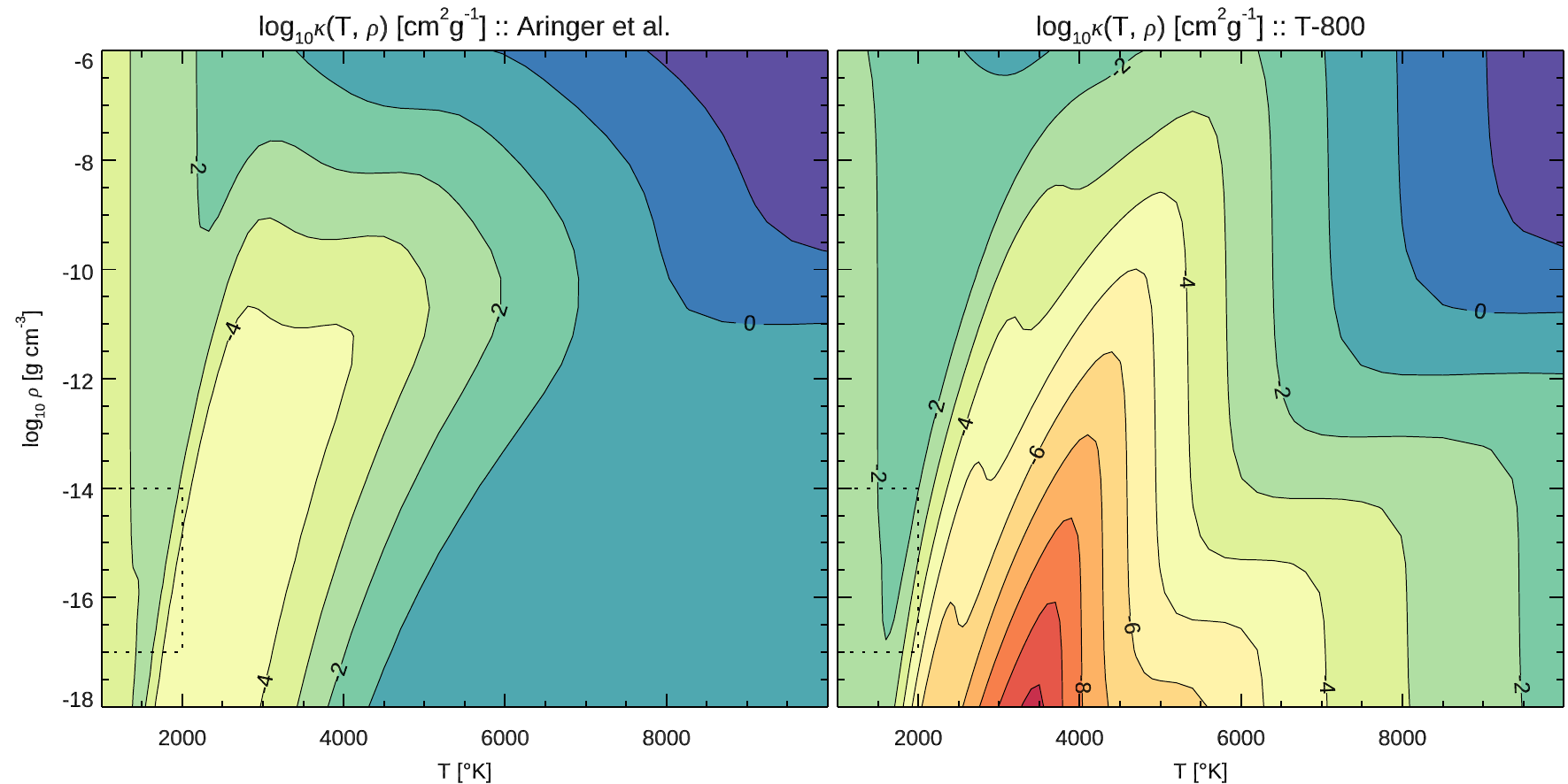}
    \caption{Total gas opacity at   wavelength   closest to $\lambda=1\,\mu$m for the  \citeauthor{ArGiNo.:09} data (left panel) and our new data (right panel). The opacity is  colour-coded and is shown vs  temperature ($x$-axis) and  gas density ($y$-axis). All integers on the contour lines are negative (except 0). The temperature and density ranges are $1000\le T\le10\,000\,$K and $-18\le\log\rho\le-6\,\text{g}\,\text{cm}^{-3}$, respectively. The two panels use the same ranges and colour palette. The dotted box delimits densities and temperatures in the wind formation region of the model shown in Fig.~\ref{fig:poc_radial}.}
    \label{fig:app}
\end{figure*}

We calculated new gas opacity tables based on the  {\exomol} data (see Sect.~\ref{sec:kappagas}). The extant and new opacity datasets are compared in Fig.~\ref{fig:app} for the wavelength $\lambda=1\mu$m. The figures show a somewhat similar pattern. Discrepancies are smaller at the lowest temperatures, where the offset is about 0.7 dex (not shown). Differences are also larger at lower densities. At higher temperatures and lower densities, our new bound-bound opacities drop faster to low values. The region of typically assumed densities and temperatures is indicated in the figure. Here bound-bound opacities of molecules dominate, and our values are about one dex lower than those of \citeauthor{ArGiNo.:09}. The same low values are not seen in the opacity data of \citeauthor{ArGiNo.:09}, whose opacity values at the higher temperatures $T\ga4000\,$K are up to $10^{6}$ times higher than our values. More recent publicly available weighted grey opacities \citep[\textsc{{\AE}sopus} project of][]{MaArGiBr:22}, which  were at some point based on the opacity data we already have, include opacities of both atoms and ions at higher temperatures (Aringer, priv. comm.). In our new opacities, the data at these higher temperatures only consist of free-free and bound-free components. The opacity data of \citeauthor{ArGiNo.:09} show values that appear to be more constant at higher temperatures and lower densities; their data show less structure than our new opacities.

With our new opacities, the stellar wind calculations are not hampered by the same kind of noise at $T\simeq2000\,$K described above. The extant opacity tables only allow us to speculate on why these data give rise to noise in the temperature when we use them to attempt to calculate a stellar wind.

\end{appendix}

\end{document}